\documentclass[twocolumns]{aa520}
\usepackage{graphicx}
\usepackage{txfonts}
\begin{document}
   \title{IMF Effects on the Colour Evolution of Disk Galaxies}

   \author{P. Westera
          \inst{1}
          \and
          M. Samland
          \inst{1}
          \and
          S. J. Kautsch
          \inst{1}
          \and
          R. Buser
          \inst{1}
          \and
          K. Ammon
          \inst{1}
          }

   \offprints{P. Westera}

   \institute{Astronomical Institute. Department of Physics and Astronomy,
             University of Basel, Venusstr. 7,
             CH-4102 Binningen, Switzerland
             \email {westera@astro.unibas.ch, buser@astro.unibas.ch}
             }

   \date{Received; Accepted}

   \abstract
             {}
             {In this work, we want to find out if the IMF can be determined
             from colour images, integrated colours,
             or mass-to-light ratios, especially at high
             redshift, where galaxies cannot be resolved into individual stars,
             which would enable us to investigate dependencies of the IMF
             on cosmological epoch.}
             {We use chemo-dynamical models to investigate the
             influence of the Initial Mass Function (IMF) on the evolution of
             a Milky Way-type disk galaxy, in particular of
             its colours.}
             {We find that the effect of the IMF on the internal gas
             absorption is larger than its effect on the light from the
             stellar content.
             However, the two effects work in the opposite sense: An IMF with
             more high mass stars leads to brighter and bluer star-light,
             but also to more interstellar dust and thus to more absorption,
             causing a kind of ``IMF degeneracy''.
             The most likely wavelength region in which to detect IMF effects
             is the infrared (i. e., $JHK$).
             We also provide photometric absorption and inclination
             corrections in the SDSS $ugriz$ and the HST WFPC2 and NICMOS
             systems.}
             {}


   \maketitle
%

\section{Introduction}

 The determination of the Initial Mass Function (IMF) of stellar populations
 and the detection of its possible variations are long lasting questions in
 astronomy.
 The first IMF, a single-slope power law, was published by Salpeter
 \cite{salpeter} based on stars in the solar neighbourhood, and is still
 occasionally used in stellar population studies.
 However, it has been known since Miller \& Scalo's \cite{millerscalo}
 milestone paper on the subject that the IMF flattens at low masses.
 In the meantime, this finding has been confirmed in numerous works
 (Scalo \cite{scalo1}, Kroupa et al. \cite{kroupaetal}, Gould et al.
 \cite{gould}, Reid \& Gizis \cite{reidgizis}, Gould et al. \cite{gould2},
 Chabrier \cite{chabrier}, Piotto \& Zoccali \cite{piottozoccali},
 Zoccali et al. \cite{zoccali}, and others).
 Recently, Chabrier \cite{chabrier02} also found indications of a turn-over
 in the brown dwarfs regime.
 For recent reviews see Kroupa \cite{kroupa02}, Chabrier
 \cite{chabrier03}. \\
 An indication of the importance of the subject is the number
 of IMFs produced over the years: Salpeter \cite{salpeter}, Miller \&
 Scalo \cite{millerscalo}, Lequeux \cite{lequeux}, Kennicutt \cite{kennicutt},
 Scalo \cite{scalo1}, Ferraro et al. \cite{ferraro}, Piotto et al.
 \cite{piotto}, Scalo \cite{scalo2}, Carigi et al. \cite{carigi}, Kroupa
 \cite{kroupa01} (universal and present day IMFs), Chabrier \cite{chabrier},
 and others. \\
 Another important issue are variations of the IMF with the star forming
 conditions (pressure, density, metallicity of the forming cloud, etc.).
 Although such variations are predicted, only little evidence of them has
 been found so far (Kroupa \cite{kroupa01}).
 We expect especially at high redshift to see differences from the present-day
 IMF, as the lower metallicity of the star-forming clouds is expected to
 cause higher temperatures and thereby higher average stellar masses
 (Larson \cite{larson}).
 Unfortunately, high redshift galaxies are too faint to be resolved into
 individual stars, which makes it difficult, if not impossible, to determine
 their IMFs.
 Therefore, it would be interesting to know if other, more global observables,
 such as integrated spectra or colours, can also yield some information about
 the stellar IMF.
 The IMF influences the light of a galaxy not only directly through the
 contributions of the stars of different (birth) masses, but also indirectly
 by affecting the entire evolution.
 The fraction of high mass stars determines the gas- and ``dust''-yield
 of the partial populations and, hence, the star formation history
 (SFH) from the second stellar generation on, as well as the gas
 absorption.
 For these reasons, we hope to be able to see signatures of the IMF even
 in the integrated light of galaxies. \\
 For this purpose, Portinari et al. \cite{portinari} studied the influence
 of the IMF on the ($I$ band) mass-to-light ($M/L$) ratio of galactic disks
 (Sbc/Sc) using chemo-photometric models and adopting 6 different IMFs,
 including the Salpeter \cite{salpeter} and Kroupa \cite{kroupa} IMFs,
 which is interesting in connection with this work, because
 in the present work, we also compare models using the Salpeter IMF and
 a more recent IMF by Kroupa.
 For each IMF, they calculate chemical evolution models with infall,
 metallicity gradients, and SFHs representative of late-type spiral disks
 (but not varying with the IMF).
 They find that so-called ``bottom-light'' IMFs (i. e., with less low-mass
 stars than Salpeter) yield low $M/L$ ratios ($\sim 0.7 - 1$), in
 agreement with various dynamical arguments and cosmological simulations.
 However, they calculate only stellar $M/L$ ratios without taking into
 account gas absorption. \\
 In this work, we use fully consistent 3-dimensional chemo-dynamical models
 by M. Samland to calculate the evolution of two galaxies, with the same
 boundary conditions (cosmology, gas infall history, etc.), but adopting
 different IMFs: the Salpeter and the Kroupa \cite{kroupa01}
 ``universal'' IMFs. Although these are not the
 most state-of-the-art IMFs available, they were chosen because of their
 clear differences in their low-to-high-mass stars ratios, so
 any IMF-induced differences should appear clearly.
 The spectra and (Hubble Space Telescope (HST), SLOAN Digital Sky Survey
 (SDSS) and Washington) colour images and integrated
 colours were then calculated using the same method as in
 Westera et al. \cite{westera}.
 An important advantage of our programme is that we can disentangle
 different effects on the spectral properties of a model galaxy, such as
 of internal absorption, by artificially blinding these contributions out,
 and then recalculating the spectral properties. \\
 The outline of this paper is as follows: Section~\ref{models} describes
 the physics and properties of the two chemo-dynamical models, and in
 Section~\ref{colours}, it is explained how the spectral properties
 of these models were calculated.
 Section~\ref{results} contains the results, and a comparison of the model
 colours with SDSS data, which were extracted in the way described in
 Section~\ref{sex}.
 In Section~\ref{summary}, we draw some conclusions and give a brief summary.

\section{The chemo-dynamical models}
\label{models}
\begin{figure}
\centering
\includegraphics[width=\columnwidth]{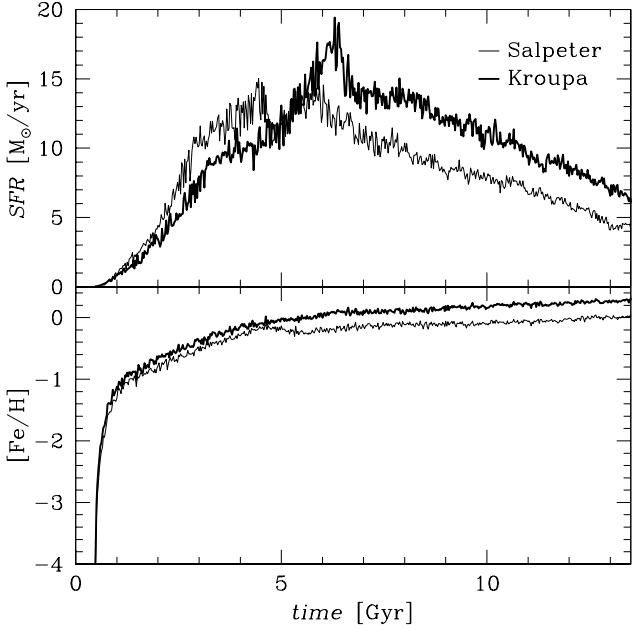}
\caption{Star formation histories of the Salpeter (thin) and Kroupa (thick) models.
                Top panel: star formation rates, bottom panel: average
                star formation metallicities.}
\label{SFH}
\end{figure}
\begin{figure}
\centering
\includegraphics[width=\columnwidth]{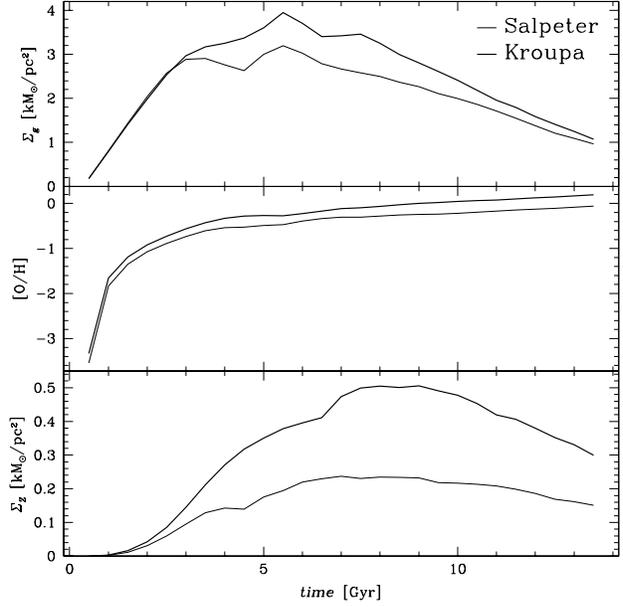}
\caption{Average gas column densities (top panel), gas metallicities (middle
                panel), and ``dust'' gas column densities, that is
                $Z_g*\Sigma_g$, of the Salpeter
                (thin) and Kroupa (thick) models.}
\label{FeH}
\end{figure}
\begin{figure*}
\centering
\includegraphics[width=\textwidth]{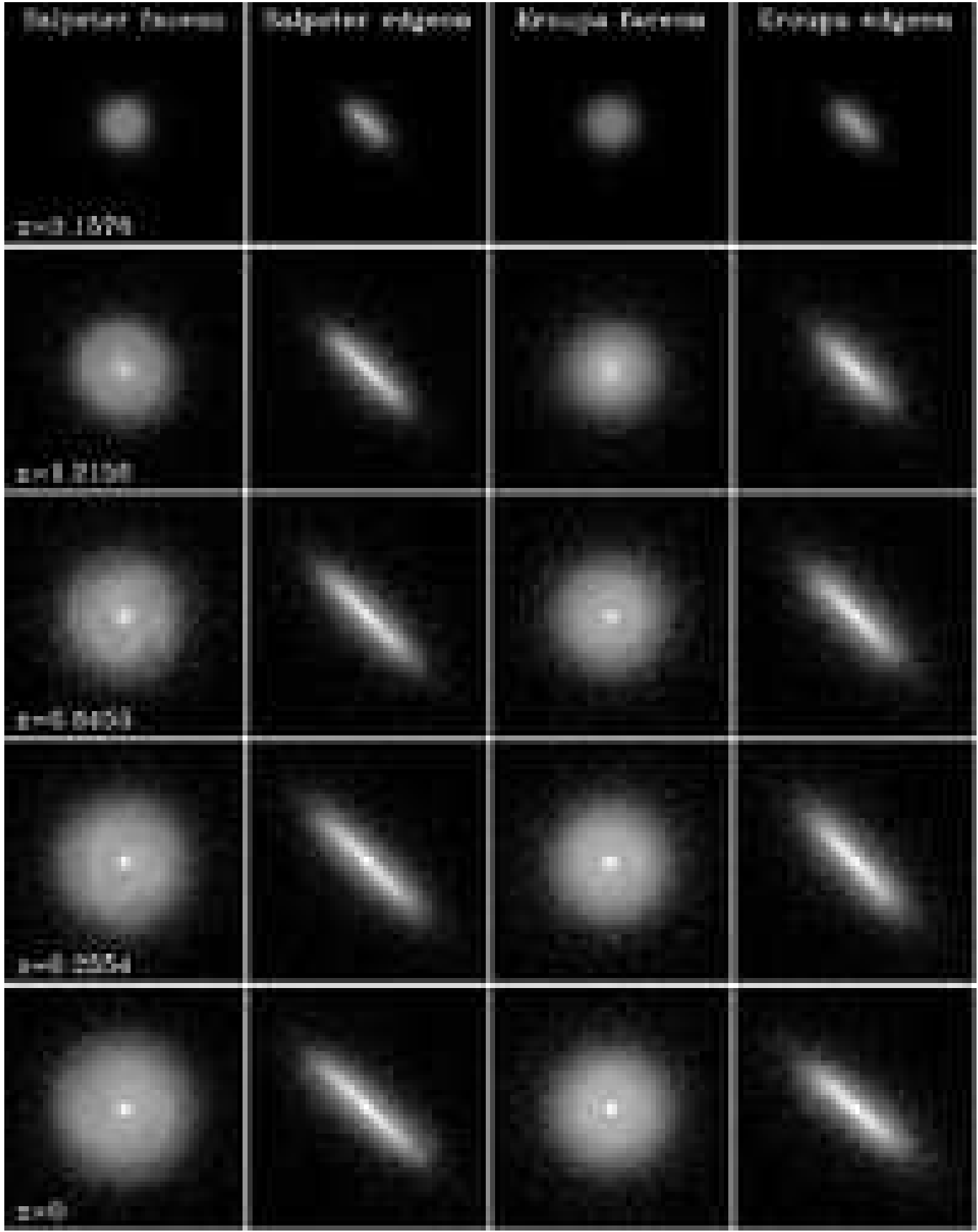}
\caption{Stellar mass distributions of the Salpeter and Kroupa models at
                different redshifts/ages (the redshifts correspond to
                ages of 2, 5, 7.5, 10.5, and 13.5 Gyr) seen face-on and
                edge-on. The colour scale is logarithmic and covers four
                orders of magnitudes. The images show an area of
                40x40 kpc.}
\label{massimage}
\end{figure*}
\begin{figure*}
\centering
\includegraphics[width=\textwidth]{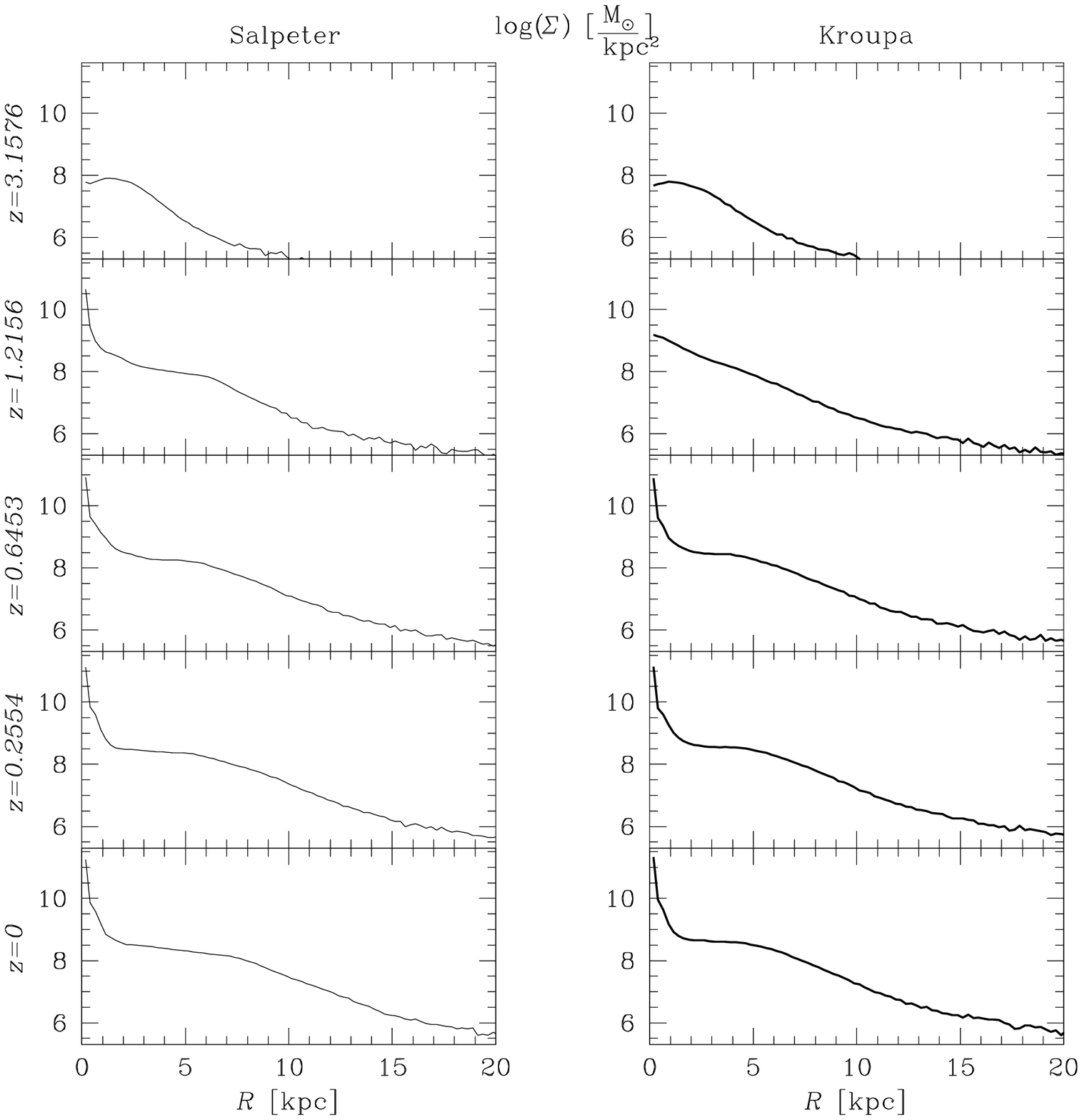}
\caption{Stellar (face-on) mass profiles of the Salpeter and Kroupa models at
                different redshifts/ages. The profiles correspond to the
                first and third columns of Fig~\ref{massimage}.}
\label{massprofiles}
\end{figure*}
%
   The models simulate the formation of a $8 \cdot 10^{11} M_{\odot}$
   galaxy leaving all parameters, except for the IMF, the same for both
   models.
   The models are for a Milky Way-type galaxy, but we expect our
   results (IMF effects) to be similar for other disk galaxies, as
   the properties, which are important for the colour evolution (star
   formation history, gas feedback and enrichment, etc.) probably
   depend in a similar way on the IMF for different types of disk
   galaxies.

   The two models have different IMFs implemented, a Salpeter
   \cite{salpeter} IMF, which is a one-segment potential law with an
   exponent $\alpha=2.35$ in the mass range from 0.1 to 50 $M_{\odot}$,
   and a Kroupa ``universal'' (also called ``standard'' or ``canonical'')
   IMF (Eq. 2 of Kroupa \cite{kroupa01}), a two-segment law with
   $\alpha=1.3$ from 0.08 to 0.5 $M_{\odot}$ and $\alpha=2.3$ from
   0.5 to 50 $M_{\odot}$, respectively, in order to investigate the
   influence of the IMF shape on the formation processes of a disk galaxy.
   The main difference between the two IMFs lies in the high-to-low mass
   stars ratio, in the sense that a population with a Kroupa IMF has more
   high-mass stars than a population of the same mass, but with a Salpeter
   IMF.

   The 3-dimensional chemo-dynamical models are of the same type as those
   described in Samland \& Gerhard \cite{samland}; so here, we only summarise
   very briefly the main properties, but take a more detailed look at the few
   quantities that will become interesting for the interpretation in
   Subsection~\ref{results}:
   the (stellar) mass surface density, the stellar particle ages, the stellar
   metallicities, and the gas density and metallicity.

   The models take into account initial cosmological and environmental
   conditions, but also internal feedback processes, such as heating by
   supernovae, dissipation, radiative cooling, nucleosynthesis, and
   in- and outflows.
   Since they are fully self-consistent models, they include dark matter,
   stars and the different phases of the interstellar medium (ISM), as well as
   the processes (``chemistry'') which connect the ISM and the stars. \\
   More quantitatively, the models assume a flat Universe with the following
   cosmological parameters:
   $H_{0} = 70 \frac{km}{s \cdot Mpc}$, $\Omega_{0} = 0.3$,
   $\Omega_{\Lambda} = 0.7$, and $\frac{M_{Bary}}{M_{Dark}} = \frac{1}{8}$.
   The total spin parameter of the model galaxies was chosen to be
   $\lambda = 0.05$ (Barnes \& Efstathiou \cite{barnes}), and the angular
   momentum distribution was calculated according to Bullock et al.
   \cite{bullock} using $\mu = 10$.
   We follow the evolution from $z=9.5$ (corresponding in this cosmology to an
   age of the Universe of 0.5 Gyr) until $z=0$ (13.5 Gyr). \\
   The models are characterised by a slowly growing dark halo and a
   continuous gas infall following the universal mass accretion histories
   found by van den Bosch \cite{vdbosch} and Wechsler et al. \cite{wechsler}
   using a formation redshift $z_{Formation}$ of 4.0 and a total mass $M$
   of $8 \cdot 10^{11} M_{\odot}$.
   As the gas infall continues until the present day,
   we expect for both models a mixture of stellar populations of
   many different ages.

   However, the two models do not show the same SFH.
   As the Kroupa IMF features more high mass stars, star formation will
   result in a stronger heating by stellar radiation, more
   stellar wind, more feedback from supernovae I and II and
   thus a higher mass return and metal yield.
   As a result, the Kroupa model has a lower SFR than the Salpeter model
   for the first $\sim 5$ Gyr, due to the heating from the stellar winds
   from the first generations.
   Afterwards, that is after $\sim 5.5$ Gyr, the Kroupa model has a higher
   SFR due to the larger available amount of gas (as seen in Fig.~\ref{FeH},
   top panel).
   After around 7.5 Gyr, this higher SFR has compensated for the lower SFR
   in the beginning, so from that point on the total stellar mass is higher
   in the Kroupa model galaxy.
   In the end (at 13.5 Gyr or redshift 0), the total stellar mass
   amounts to $1.07 \cdot 10^{11} M_{\odot}$ in the Salpeter model and to
   $1.25 \cdot 10^{11} M_{\odot}$ in the Kroupa model.
   The two SFRs can be studied in detail in Fig.~\ref{SFH} (top panel),
   from which can also be seen that the SFR remains significant until the
   present epoch, as expected from the gas infall history.
   The bottom panel shows the average star formation metallicity.

   The average gas
   metallicities ${\rm [O/H]}$ of the models (shown in Fig.~\ref{FeH},
   middle panel) increase most steeply during the phases of maximum
   star formation. They start at ${\rm [O/H]} \simeq -4$, and reach
   their present values of $\sim -0.1$ dex or $\sim +0.2$ dex
   at $z \simeq 1$.
   This higher gas metallicity of the Kroupa model, combined with the
   higher gas density (top panel of Fig.~\ref{FeH}), causes the Kroupa
   model galaxy to contain about twice as much metals in the interstellar
   matter (ISM), or ``dust'', as the Salpeter galaxy, which can be seen
   in the bottom panel of Fig.~\ref{FeH}.

   The output quantities of interest (which are the input quantities for the
   programme which calculates the spectral properties) are the following, at
   each time step: a number of stellar particles, each with its spatial position,
   initial mass, age, and metallicity, as well as the gas density and
   metallicity on a 3-dimensional grid covering the galaxy out to where
   the gas density is negligible (100 kpc), as a function of time.

   Fig.~\ref{massimage} shows the stellar mass distributions of the
   two models projected face-on and edge-on.
   Both models result in disk galaxies with weak spiral arms, whereas
   the Salpeter model produces a bulge 2 Gyrs sooner (at $\sim 4$ Gyr) than
   the Kroupa model, which can be seen in the second row of
   Fig.~\ref{massimage} ($z=1.2156$, which corresponds to 5 Gyr).
   In the Kroupa model (right two panels), the bulge is not yet present at
   this epoch, but in the Salpeter model (left two panels), it is.
   This can also be seen in the profiles (Fig.~\ref{massprofiles}, second
   row).
   This delay in bulge formation in the Kroupa model is probably also due
   to stellar winds.
   As soon as the bulge appears in either model, we also see a
   plateau in the mass profiles at around 2 to 5 kpc. This is due
   to stellar winds from the bulge, which push out the gas from the inner
   disk (at 2 kpc) to a distance of 4 kpc, where Star Formation then
   takes place.
   It is not bar-induced as in the more massive galaxy studied by
   Samland \& Gerhard \cite{samland}.
   The fact, that, in Fig.~\ref{massimage}, the Kroupa model galaxy seems
   to have a thicker disk than the Salpeter one, is just a by-eye
   impression.
   We calculated the thick and thin disk scaleheights for both models
   as functions of time, but found no significant differences between
   the models.

\section{From theoretical quantities to colours and spectra}
\label{colours}
\begin{figure*}
\centering
\includegraphics{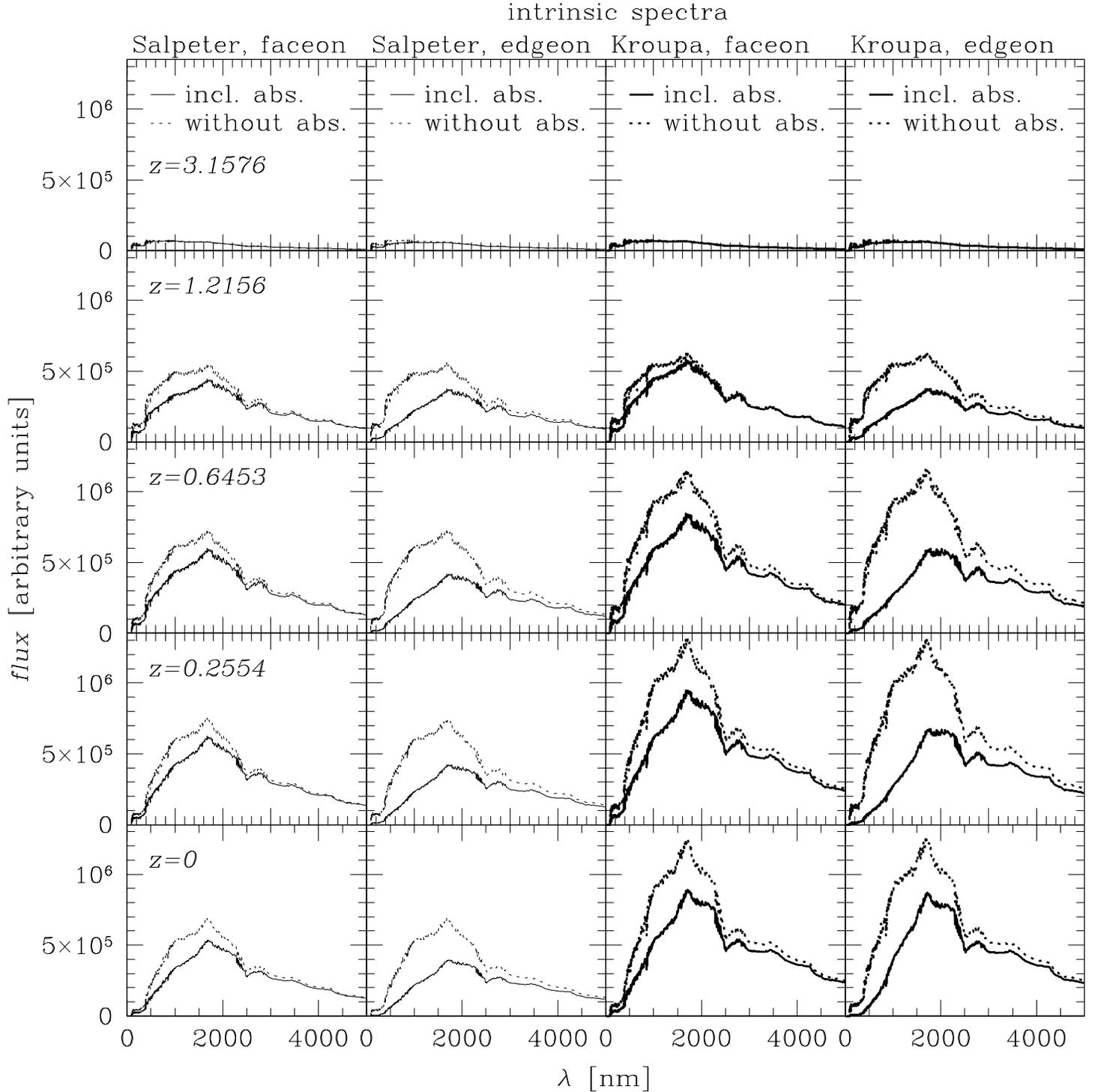}
\caption{Intrinsic spectra of the Salpeter (left two columns) - and
       Kroupa (right two columns) models at five different epochs
       (different rows), both face-on (columns 1 and 3) and edge-on
       (columns 2 and 4), and both without (dotted) and including
       (solid) absorption.}
\label{spectra}
\end{figure*}

   To derive 2-dimensional colour images (HST (WFPC2 and NICMOS), SDSS $ugriz$,
   Washington $CNT1T2$, and other photometric systems)
   from the star and gas distributions of the
   galaxy models, we proceeded in the following way: \\
   First, two libraries of simple stellar population (SSP) spectra were
   produced: one with a Salpeter IMF from 0.1 to 50 $M_{\odot}$, and one
   with a Kroupa IMF from 0.08 to 50 $M_{\odot}$, in accordance with the
   galaxy models.
   With the Bruzual and Charlot 2000 Galaxy Isochrone Spectral Synthesis
   Evolution Library (GISSEL) code (Charlot \& Bruzual \cite{charlot_91},
   Bruzual \& Charlot \cite{bruzual_93}, Bruzual \& Charlot \cite{bruzual_03}),
   integrated spectra (ISEDs) of populations were calculated for
   a grid of population parameters consisting of 7 metallicities
   (${\rm [Fe/H]} = -2.252$, --1.65, --0.65, --0.35, 0.09, 0.447,
   and 0.748) and 221 SSP ages ranging from 0 to 20 Gyr.
   As input, we used Padova 1994 isochrones  (Fagotto et al. \cite{fagotto2},
   Girardi et al. \cite{girardi}).
   There exist more recent versions of the Padova isochrones, Padova 2000
   (Girardi et al. \cite{girardi_00}), but there is some doubt as to
   whether these newer tracks produce better agreement with observed galaxy
   colours than the Padova 1994 models (Bruzual \& Charlot \cite{bruzual_03}).
   Furthermore, the Padova 1994 isochrones cover a wider range of metallicities.
   The spectral library used was the BaSeL 3.1 ''WLBC 99''
   (Westera \cite{diss}, Westera et al. \cite{paperiii})
   stellar library.
   The spectra of this ISED library contain fluxes at 1221 wavelengths
   from 9.1 nm to 160 $\mu$m, comfortably covering the entire range where
   galaxy radiation from stars is significant. 
   The GISSEL software also has a higher resolution stellar library
   implemented, STELIB, which has a resolution high enough to study spectral
   (absorption) lines ($1 \AA$ in the relevant wavelength range), but, with
   6900 flux points per spectrum, these spectra proved too large in terms of
   memory and CPU time to be included in our programme. \\
   After choosing (through three angles) the viewing direction with respect to
   the galaxy principal plane, and the size (up to $320 \times 320$ pixels) and
   resolution for the ``virtual CCD camera'', the stellar particles are grouped into
   pixels. For each stellar particle, the spectrum is (geometrically,
   flux point by flux point) interpolated from the ISED library.
   For metallicities lower than the range covered by the library,
   the spectra for the lowest metallicity (${\rm [Fe/H]} = -2.252$)
   were used. This should not pose any problems, as trends of spectral
   properties with metallicity are expected to become weak below
   ${\rm [Fe/H]} = -2.0$, and these lowest-metallicity stellar particles become
   negligible in number very soon.
   For SSPs of 50 Myr and younger, we added nebular emission to the spectra
   in the same way as described in Leitherer et al. \cite{leitherer}, and
   accordingly removed the flux below 912 \AA.
   On the other hand, the emission of HII regions is not implemented.
   The inclusion of HII regions, as well as planetary nebulae and
   supernovae, will be one of the next steps in improving the
   programme. \\
   Then, the spectra were reddened as follows, using the gas density
   and metallicity in the model to trace the three-dimensional
   distribution of dust: For each stellar particle,
   the metallicity-weighted gas density was integrated along the line of sight
   to derive the absorption coefficient $A_{V}$ according to Quillen \& Yukita
   \cite{quillen_01}:
   \begin{equation}
   A_{V}=\frac{1}{50\frac{\mathrm{M}_{\sun}}{\mathrm{pc}^{2}}}
   \int_{LOS}\rho_{g}(r)\left( \frac{Z(r)}{Z_{\sun}}\right) dr\ .
   \end{equation}
   The spectrum of the stellar particle was then reddened
   using the extinction law of Fluks et al. \cite{fluks_94}. \\
   All the spectra of stellar particles from the same pixel were added
   up to give the integrated absolute spectrum of the pixel, which was
   then redshifted and dimmed using the redshift $z$ from the models,
   and calculating the distance modulus $m-M$ according to Carroll et al.
   \cite{carroll_92}:
   \begin{eqnarray}
   \label{mM}
     m-M(z)= \nonumber\\
     5\log \left( \frac{c}{H_0}
   (1+z)\int_{0}^{z}[(1+z')^2(1+\Omega_Mz') \right.\nonumber\\
    \left.  -z'(2+z')\Omega_\Lambda]^{-1/2}dz' \right) + 25\ .
   \end{eqnarray}
   We then corrected the spectra for Lyman line blanketing and Lyman
   continuum absorption by absorption systems at cosmological distances
   using the formulae given by Madau \cite{madau_95} for QSO absorption
   systems.
   Finally, apparent HST (WFPC2 and NICMOS), SDSS $ugriz$, Washington $CNT1T2$
   colours and magnitudes were calculated for each pixel through synthetic
   photometry.
   Other photometric systems, i. e. Johnson-Cousins $UBVRIJHKLM$,
   Str\"{o}mgren $ubvy$, Kron $RI$,
   are also implemented in the programme, but we limited our study to the
   above-mentioned systems for memory - and CPU time reasons.
   Including more systems is unlikely to yield further discoveries, since
   the filter bands of those systems lie in the same wavelength range as the
   ones of the systems we used, and will thus most probably show the same
   behaviour with IMF (and other) variations.
   Furthermore, the HST and SDSS systems seemed most likely to allow
   extensive comparison with observational data. \\
   At the same
   time, the absolute (rest frame) spectra and the apparent spectra of
   all the pixels were added up to derive the absolute and apparent
   integrated spectra of the galaxy.
   Examples of such integrated (intrinsic, that is unredshifted and all
   with the same distance modulus) spectra are shown in Fig.~\ref{spectra}.
   The metallicity-dependent distribution of the stars and the spatially
   resolved treatment of the gas absorption are the most important for
   the spectra and colours.\\
   On these integrated spectra, synthetic photometry was performed,
   too. For computer memory reasons, the spectra of individual pixels
   or stellar particles were not stored, so the final output quantities
   of the programme are:
   \begin{enumerate}
    \item a 2-dimensional colour image of the model galaxy, including
    the effect of internal absorption in intrinsic magnitudes of
    up to $320 \times 320$ pixels, as seen from a freely chosen angle,
    \item the same image in apparent (redshifted and corrected for
    the distance modulus and Lyman line blanketing) magnitudes,
    \item the integrated intrinsic spectrum of the entire galaxy plus
    integrated intrinsic colours and absolute magnitudes,
    \item the integrated apparent spectrum of the entire galaxy plus
    integrated redshifted colours and apparent magnitudes.
   \end{enumerate}
   Our programme also includes the possibility to account for Galactic
   foreground reddening. But since this option only makes sense for specific
   applications, where the foreground reddening is known, it was not used
   in this work.\\
   These quantities were calculated for both the Salpeter IMF and the
   Kroupa IMF models, at ages from $0.5$~Gyr (corresponding to $z=9.5$,
   or $0.3$~Gyr after the beginning of the simulation) to
   $13.5$~Gyr (the present day) in steps of $0.5$~Gyr, and from three
   different directions: face-on, inclined by $60^{\circ}$,
   henceforth called the diagonal view, and edge-on.
   The size of a pixel was chosen to be $0.25$~kpc.
   Higher resolution would make no sense, as the galaxy model has a
   precision of only $0.37$~kpc.
   The entire ``camera'' was chosen $320 \times 320$ pixels wide, thus
   representing a field of view of $80 \times 80$~kpc.\\
   To identify absorption effects, the same photometric
   properties were calculated for both models without internal
   absorption. Thus, the differences between the regular models and
   these ones should reflect absorption effects, or the error
   in models that do not include internal absorption. These models
   will be called the absorptionless models, and will be used in
   Section~\ref{results}.

\section{Data Extraction}
\label{sex}

In order to test our models, we compare them to galaxy data from
the Sloan Digital Sky Survey (SDSS) (York et al. \cite{yor00}), which
offers a large sample of galaxies.
The current volume of the SDSS is the Data Release 4 (DR4) which covers in
the imaging mode about 180 million unique objects in an area of 6670 square
degrees and 849,920 spectra within 4783 square degrees
(Adelman-McCarthy et al. \cite{adel05}). 
We use the SDSS Batch Query
Services\footnote{http://casjobs.sdss.org/CasJobs/default.aspx}  
on the DR4 Galaxy Table View and SpecObj Table View. This web interface allows
to perform queries on the available SDSS archives using the Structured Query
Language (SQL). The Galaxy Table View contains optical parameters of all
galaxies at the time of the data release. The spectral properties of the
galaxies are given in the SpecObj Table View.
We remove all objects which are flagged with
one or more of the PhotoFlags as given in the Galaxy Table View: blended
(object had multiple peaks detected within it); edge (object
is too close to edge of frame of the survey); saturated (object contains
saturated pixels); ellipfaint (not measured isophotal properties and
incomplete profiles). These flags allow us to reject all objects near the
survey borders and blended with spikes of nearby stars.
As our model colours differ for the face-on and edge-on view
(see Section~\ref{results}), we extract two samples, one for either
viewing direction. \\
For the edge-on sample, we adopt the query used by Kautsch et al.
\cite{wixxy06} to collect a catalog of edge-on galaxies, wherein
the axial ratio $a/b$ is chosen to be $>3$, $a$ and $b$ being the major and
minor angular isophotal axes in the $g$ band; the major axis $a$ is chosen
larger than 15 pixels (which corresponds to 5.94 arcsec) and colours
in the ranges $-0.3 < g-r < 3$ mag and $-0.3 < r-i < 3$ mag, in order to
exclude spurious objects and other artefacts.
We limit the sample to a Petrosian magnitude\footnote{the Petrosian magnitudes
are derived from the Petrosian flux using a circular aperture centered on
every object. The advantage of this method is that this allows an unbiased
measurement of a constant fraction of the total galaxy light using the
technique based on that of Petrosian \cite{petro76}. For a detailed
description of the Petrosian parameters used in the SDSS we refer to
Blanton et al. \cite{blan01} and Yasuda et al. \cite{yasu01}.}
in the $g$ band of 20. \\
For the face-on sample, $a/b$ is chosen to be smaller than 1.5, the
isophotal major axis $a$ is also $>15$ pixels, and the colours again lie
in the ranges $-0.3 < g-r < 3$ mag and $-0.3 < r-i < 3$ mag.
Here the Petrosian $g$ band magnitude is limited to 19 mag, since galaxies
seen face-on appear brighter than the same ones seen edge-on
(see Section~\ref{results}). Using these limiting magnitudes, the two samples
contain a similar number of galaxies.

The following biases affect our selection: ``Shredded galaxies,'' i. e.,
these galaxies are detected as two or more independent objects (this is found
in particular for extended objects with substructure and diameters
$\geqslant 1'$); galaxies with unusual colours caused by an AGN and/or dust.
Due to these effects we lose less than 1\% of the targets from the SDSS
database as estimated from a by-eye-inspection of randomly selected
subsamples. \\
Wrong classification can be the result of various causes:
(i) ``inverse shredding'', where objects arranged in
chains are detected as a single object; 
(ii) bars or spiral arms in faint disks being classified as edge-on galaxies.
However, we estimate that these effects affect about 2\% of the targets only.

\section{Results}
\label{results}

\begin{figure*}
\centering
\includegraphics{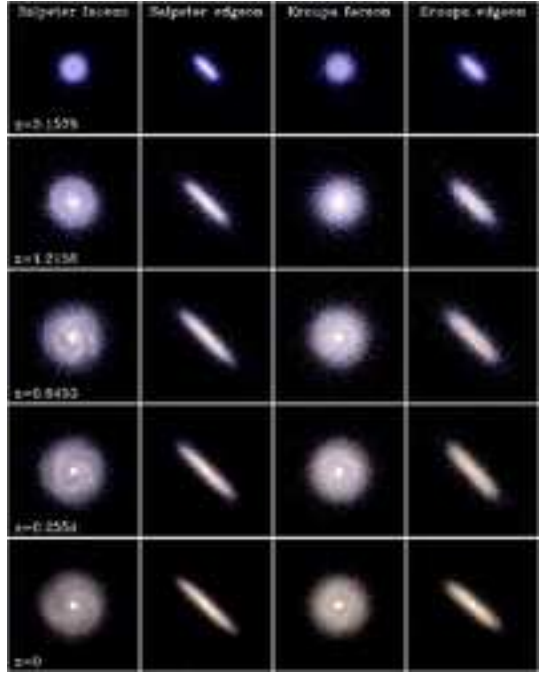}
\caption{Calculated intrinsic $urz$ images of the Salpeter
       and Kroupa models, face-on and edge-on, both in the same
       magnitude scaling, where the calculated $u$ distribution
       makes up for the blue portion of the composite image, 
       the $r$ distribution for the yellow portion, and the
       $z$ distribution for the red portion.
       The images correspond to the images in Fig~\ref{massimage}.}
\label{URZpictures}
\end{figure*}

\begin{figure*}
\centering
\includegraphics{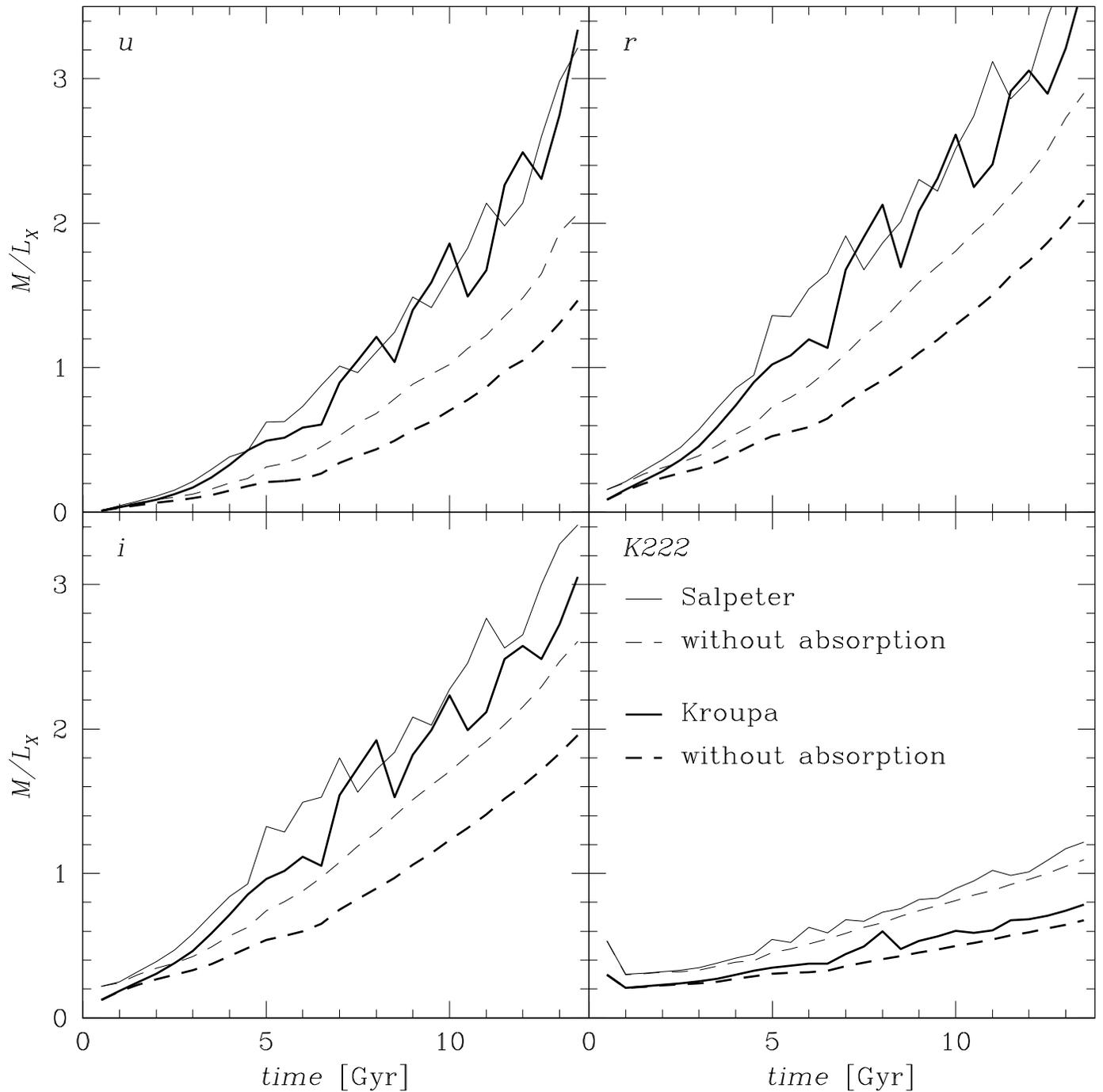}
\caption{Stellar mass-to-light ratio evolution in 
       $u$, $r$, $i$, and $K222$ of the Salpeter (thin) and the Kroupa (thick)
       models, both including (solid) and omitting (dashed) gas absorption.}
\label{MLUKsk}
\end{figure*}

\begin{figure*}
\centering
\includegraphics{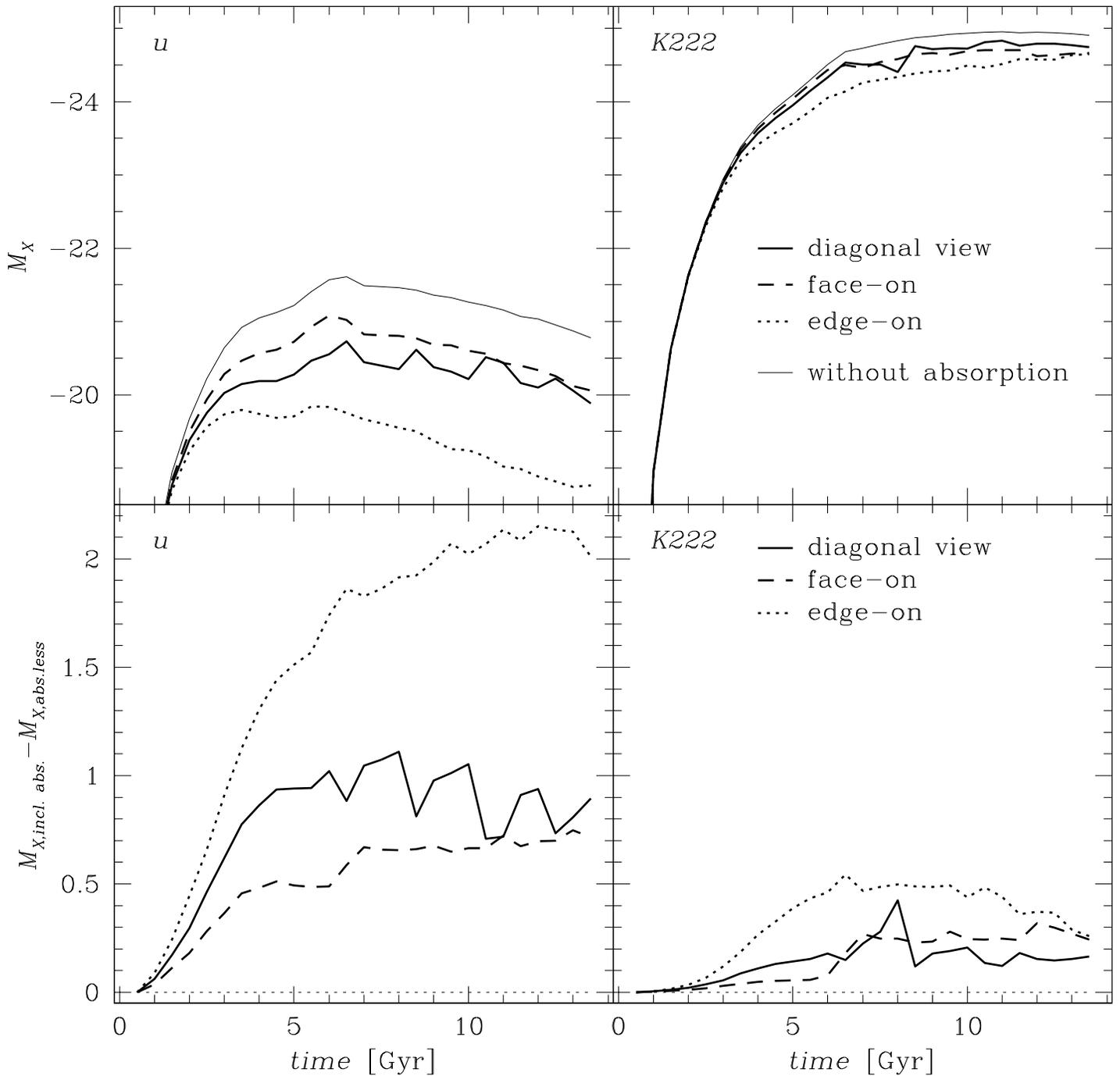}
\caption{Top panels: Absolute $u$ and $K222$ magnitude evolution of the Kroupa
       model as seen in the diagonal view (solid), face-on (dashed), and
       edge-on (dotted). The thin solid line shows the evolution of the
       absorptionless model (in the diagonal view, but it looks the same for
       the face-on and edge-on views).
       Bottom panels: Effects of the gas absorption on the absolute $u$ and
       $K222$ magnitudes of the Kroupa models (that is, the differences
       between the absorptionless and the regular models) as seen in the
       diagonal view (solid), face-on (dashed), and edge-on (dotted).
       The thin dotted lines show the zero level.}
\label{UKdb}
\end{figure*}

\begin{figure*}
\centering
\includegraphics{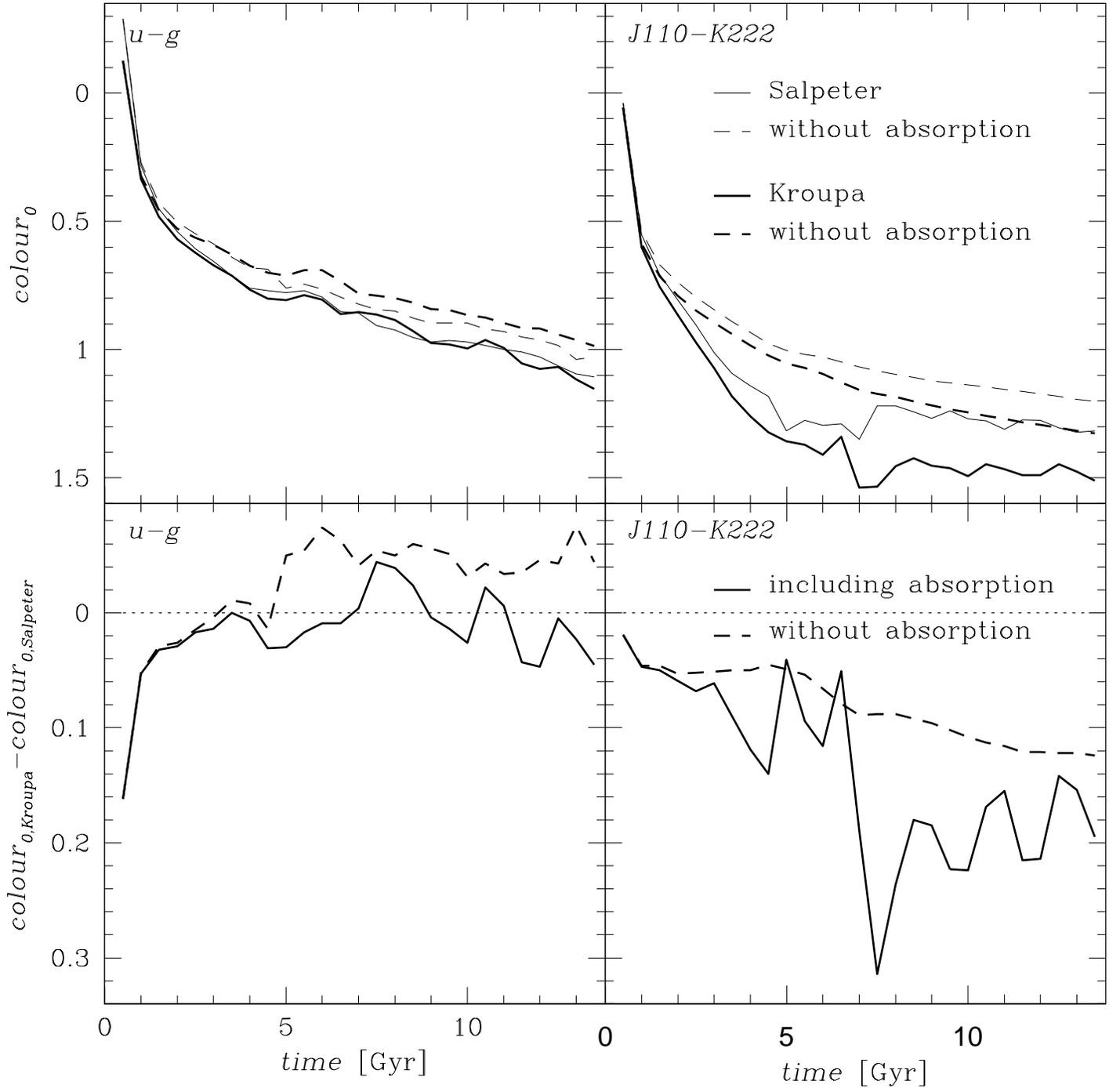}
\caption{Top panels: Intrinsic $(u-g)_0$ and $(J110-K222)_0$ colour evolution
       of the Salpeter (thin) and the Kroupa (thick) models, both including
       (solid) and omitting (dashed) gas absorption.
       Bottom panel: differences in intrinsic $(u-g)_0$ and $(J110-K222)_0$
       colours between the Salpeter and the Kroupa models, both including
       (solid) and omitting (dashed) gas absorption.
       The thin dotted line shows the zero level.}
\label{UGJKsk}
\end{figure*}

\begin{figure*}
\centering
\includegraphics{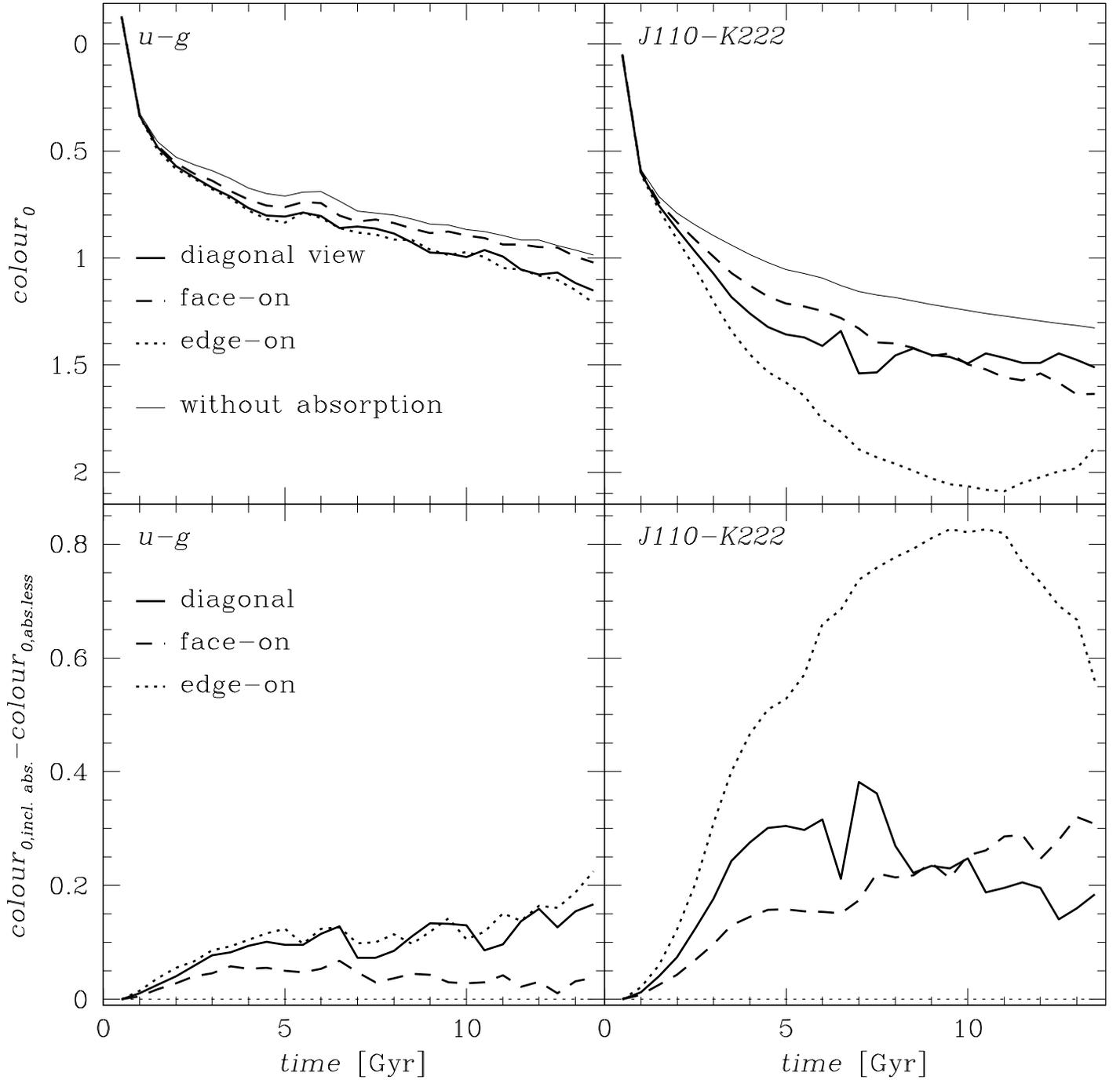}
\caption{Top panels: Intrinsic $(u-g)_0$ and $(J110-K222)_0$ colour evolution
       of the Kroupa model as seen in the diagonal view (solid), face-on
       (dashed), and edge-on (dotted).
       The thin solid line shows the evolution of the absorptionless model
       (in the diagonal view, but it looks the same for the face-on and
       edge-on views).
       Bottom panels: Absorption effects on the intrinsic $(u-g)_0$ and
       $(J110-K222)_0$ colours of the Kroupa models in each of the three
       viewing directions, that is, the differences between the absorbed
       and the absorptionless model seen in the diagonal view (solid),
       face-on (dashed), and edge-on (dotted).
       The thin dotted lines show the zero level.}
\label{UGJKdfe}
\end{figure*}

\begin{figure*}
\centering
\includegraphics{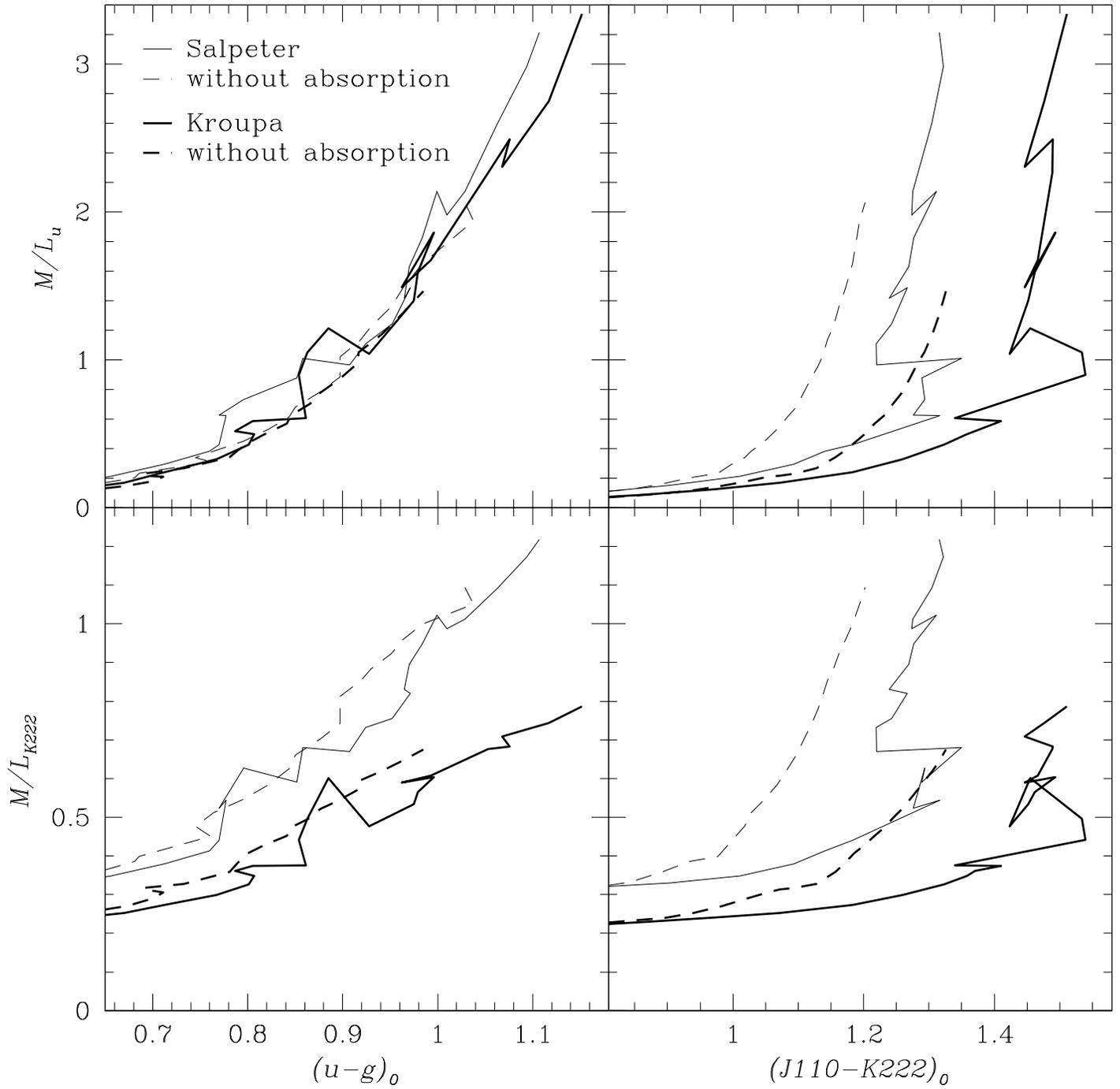}
\caption{Colour-$M/L$ relations of the two models in different
       passbands, including and omitting absorption.}
\label{colML}
\end{figure*}

\begin{figure*}
\centering
\includegraphics{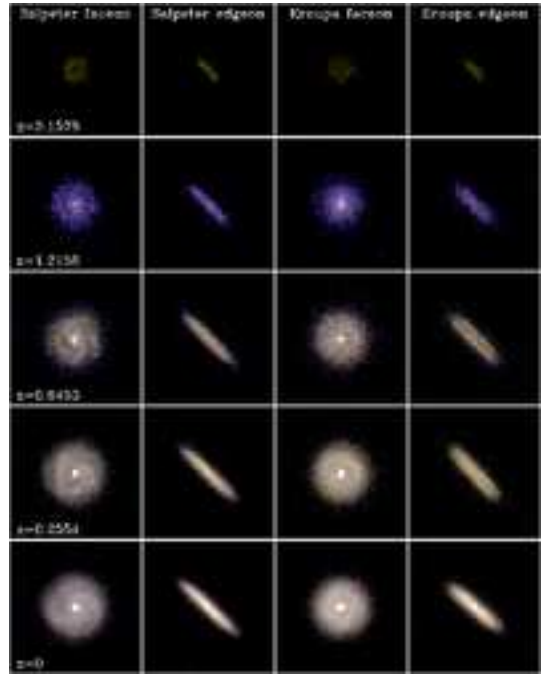}
\caption{Calculated apparent $urz$ images of the Salpeter
       and Kroupa models, face-on and edge-on, both in the same
       magnitude scaling (corrected for distance modulus),
       where the calculated $u$ distribution
       makes up for the blue portion of the composite image, 
       the $r$ distribution for the yellow portion, and the
       $z$ distribution for the red portion.
       The images correspond to the images in Fig~\ref{massimage}.}
\label{urzpictures}
\end{figure*}

\begin{figure*}
\centering
\includegraphics{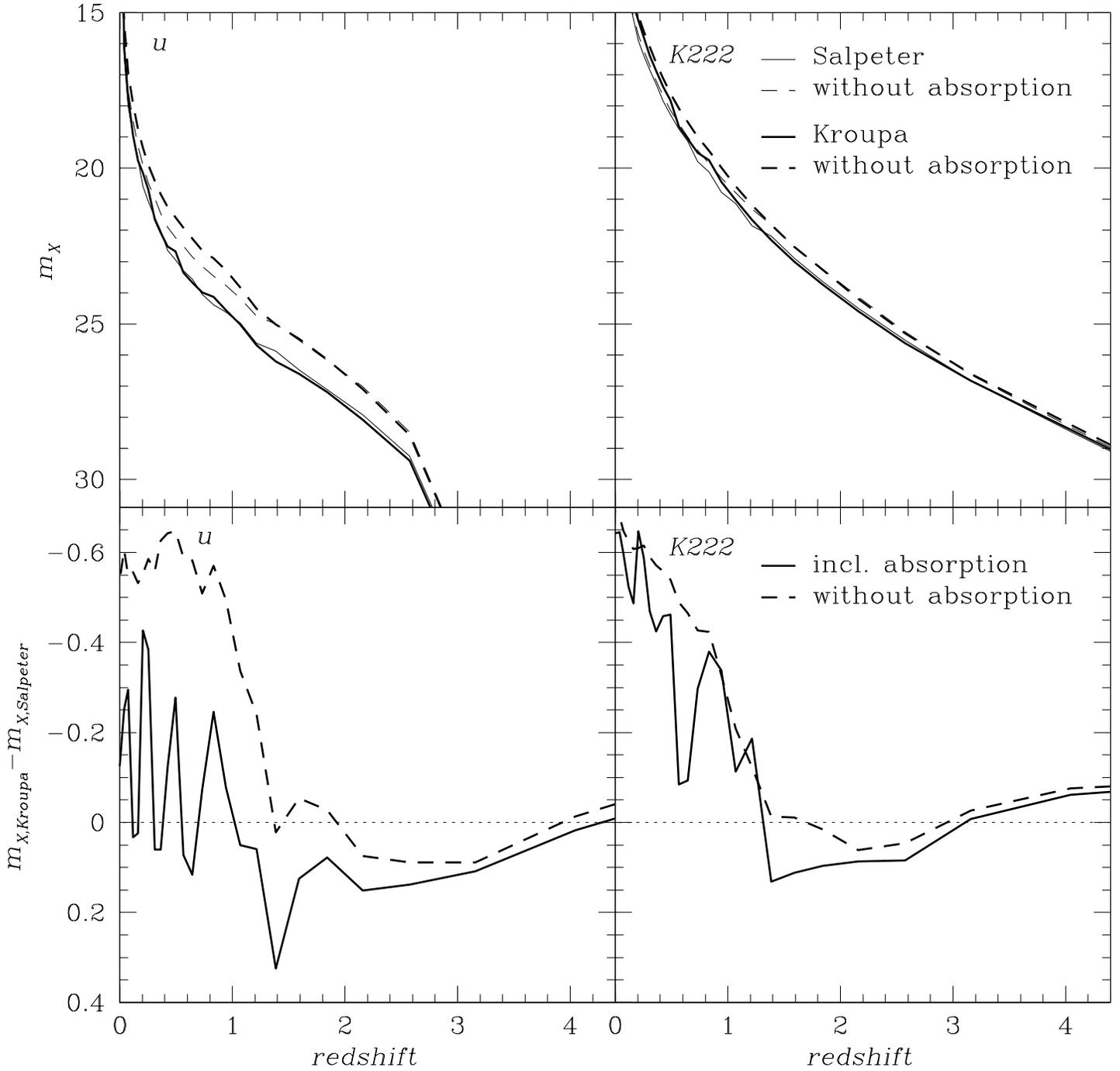}
\caption{Top panels: Apparent $u$ and $K222$ magnitude evolution of the
       Salpeter (thin) and the Kroupa (thick) models as a function of redshift,
       both including (solid) and omitting (dashed) gas absorption.
       Bottom panels: differences in $u$ and $K222$ magnitudes between the
       Salpeter and the Kroupa models, both including (solid) and omitting
       (dashed) gas absorption.
       The thin dotted lines show the zero level.}
\label{uksk}
\end{figure*}

\begin{figure*}
\centering
\includegraphics{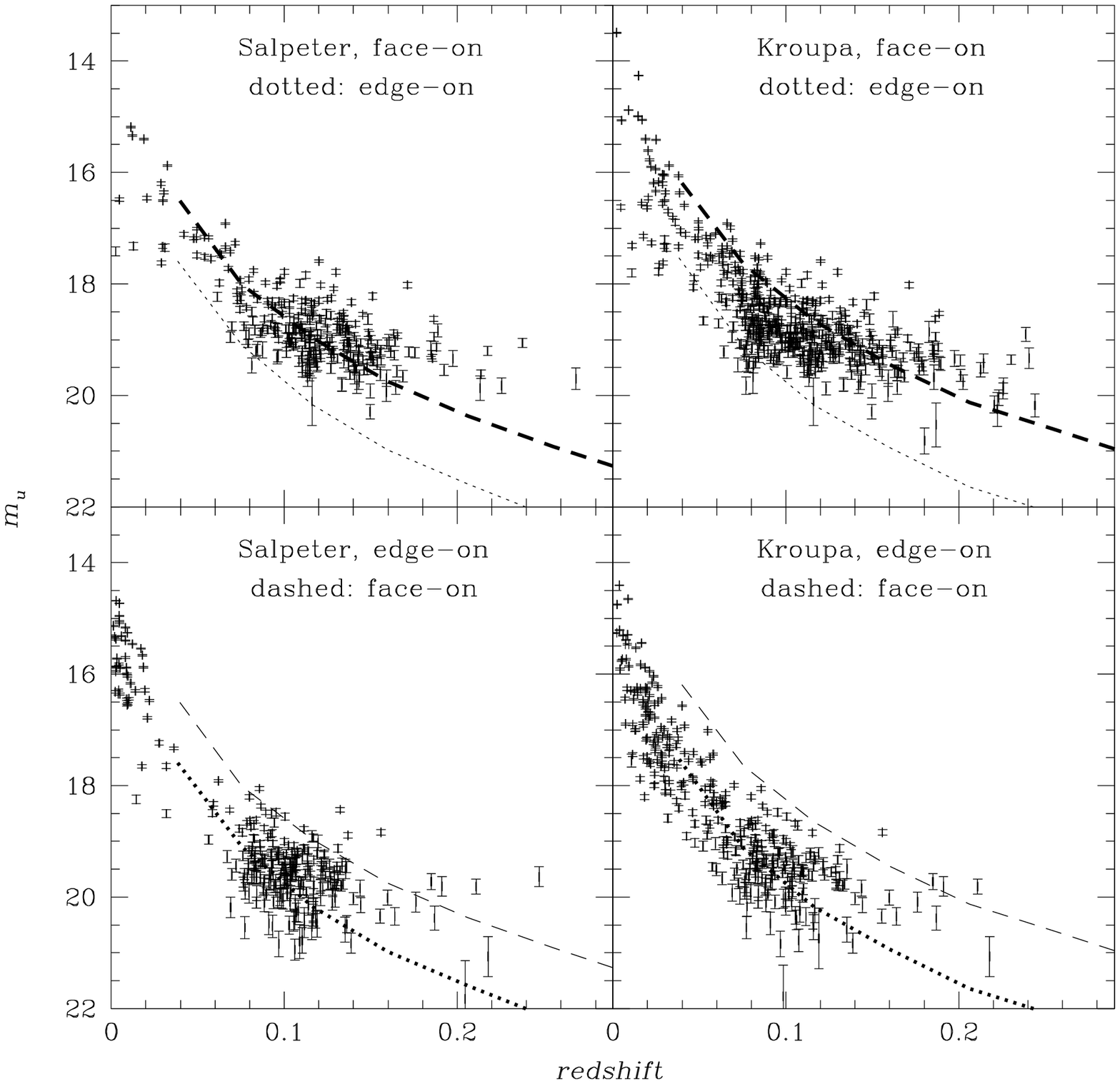}
\caption{Apparent $u$ magnitudes of the Salpeter and Kroupa models as seen
       face-on (dashed) and edge-on (dotted) as a function of redshift.
       Overlaid are the SDSS DR4 data with their observational errors of
       galaxies that have similar sizes, concentration parameters, and
       orientation as the models (for a more precise description, see the
       text).}
\label{uskS}
\end{figure*}

\begin{figure*}
\centering
\includegraphics{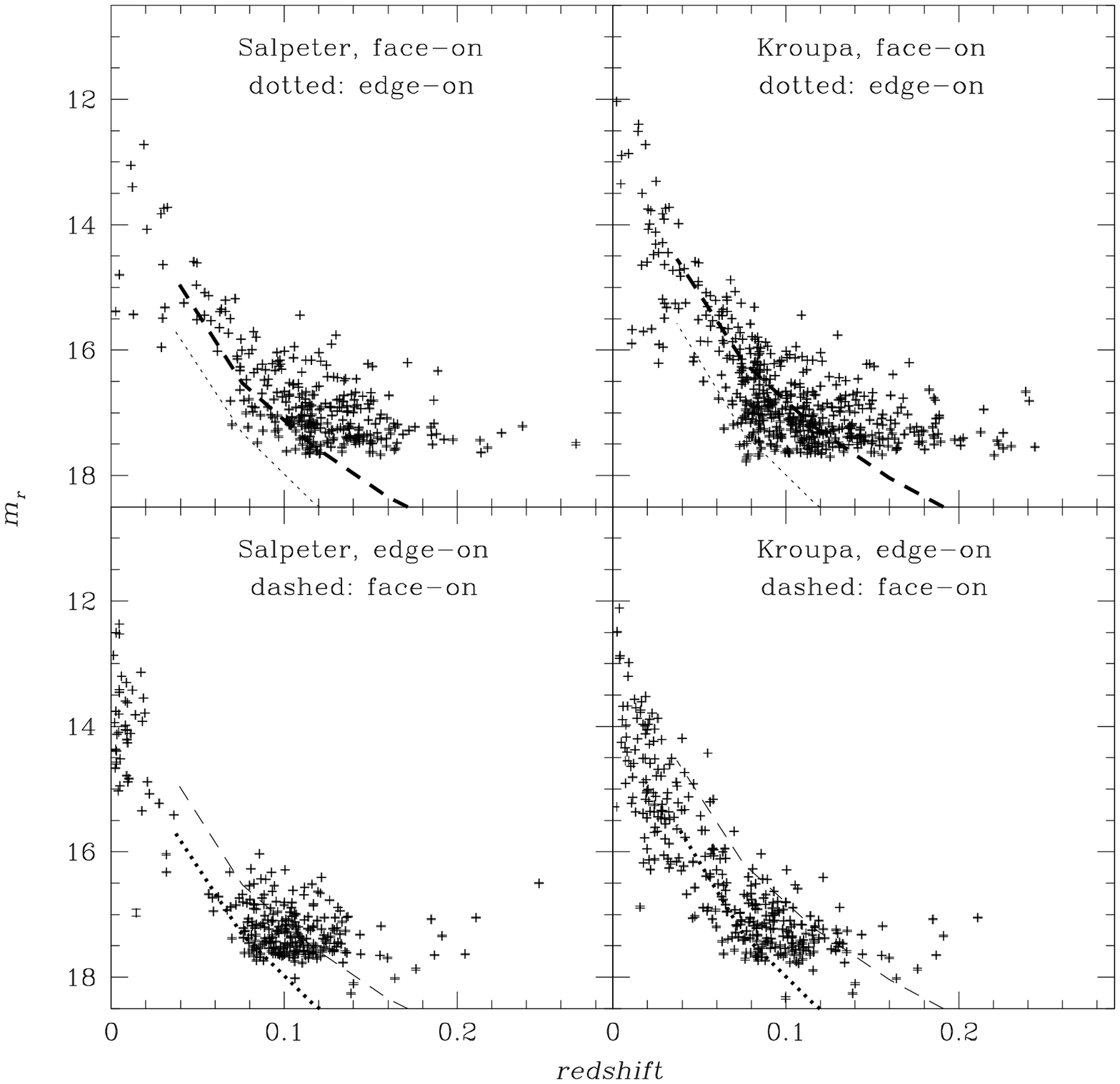}
\caption{Like Fig.~\ref{uskS}, but for the $r$ band.}
\label{rskS}
\end{figure*}
In Fig.~\ref{URZpictures}, we see the calculated intrinsic
$urz$ band images of the Salpeter and Kroupa model galaxies, both in
the same magnitude scaling.
The images confirm that the bulge forms later in the Kroupa model (at around
6 Gyr) than in the Salpeter model ($\sim$4 Gyr), as mentioned in
Section~\ref{models}. Apart from that, the images of the models look extremely
similar. In this section, we will explain why.

\subsection{Intrinsic magnitudes and mass-to-light ratios}

In Tables~\ref{salpeterintrinsicsdss} to~\ref{kroupaintrinsichst}
are listed the intrinsic bolometric, $V$ Johnson, SDSS $ugriz$,
and HST (WFPC2 and NICMOS) magnitudes of the two models integrated
over the full galaxies in the diagonal view.
We calculated the Washington magnitudes as well,
and they can also be calculated for other colour systems, such as
Johnson-Cousins $UBVRIJHKLM$, Str\"{o}mgren $ubvy$, and Kron $RI$.
The evolution of the magnitudes show some oscillations around their
mean tendencies, which are the result of 
shadowing of the bulge by dense streams of infalling molecular
gas. They only show up in the diagonal view when absorption is included.
One should keep in mind that this can be an additional source of scatter
when looking at observed magnitudes and colours of galaxies. \\
In the following, most figures concerning magnitudes will show the
SDSS $u$ band and the HST NICMOS $K222$ band, and most figures
concerning colours will show $u-g$ and $J110-K222$, which are the
bluest and reddest calculated magnitudes and colours, respectively,
and therefore are the most illustrative of the range of possible effects on
colours and magnitudes.
For other passbands and colours, the trends will usually lie between the
trends of the ones shown in the figures. \\
Tables~\ref{salpeterML} to~\ref{kroupaMLwoabs} give the stellar
mass-to-light ratios ($M/L$) of the two models both including and
omitting absorption, in bolometric light, Johnson Vj,
the SDSS $ugriz$ system, and the HST NICMOS $JHK$ system.
They are shown in Fig.~\ref{MLUKsk} in $u$, $r$, $i$, and $K222$
as a function of time since the Big Bang.
We see that, from around 6 Gyrs on, the Kroupa model has $M/L$s
about one third lower than the Salpeter model. It is $\sim 0.5$ mags
brighter, due to its higher SFR (see Fig.~\ref{SFH}).
Unfortunately, when absorption is included, this effect is partly
canceled out by the higher gas content of the Kroupa model to a level
varying between 0 ($K222$) and 0.4 mags ($u$), which is
probably undetectable.
The effect in colour bands bluewards of $i$ is, that the two models
become indistinguishable, an unfortunate coincidence we call
``IMF degeneracy''.
It is weakest in the $K222$ band, where the absorption is smaller and
the difference between the Salpeter - and Kroupa models remains around
$\sim 0.5$ mags, leaving the difference in $M/L$ unaltered,
but even this will be difficult to measure. \\
The above-mentioned absorption effects vary with the viewing angle,
as is illustrated for the Kroupa model in Fig.~\ref{UKdb} (for the
Salpeter model they look similar).
It shows the $u$ - and $K222$ evolution of the model in all angles,
as well as the absorptionless model (which has the same intrinsic
magnitudes viewed from any angle).
The lower panels show the absorption effects on these magnitudes (that is
the differences between the models with different viewing angles and their
absorptionless counterparts).
These differences translate into differences in magnitudes between
the different inclinations.
They amount to 1.5 mags in $u$, and still 0.5 mag in $K222$, which is
more than the IMF effects.
Nevertheless, it will be difficult to say something about the orientation 
of an unresolved galaxy by its intrinsic magnitude alone,
since too many things can affect the total magnitude of a real distant
galaxy, besides inclination.

\subsection{Intrinsic colours}

If IMF effects on magnitudes or Mass-to-light ratios are insignificant,
suffering from a degeneracy, does the IMF manifest itself more strongly in
the colours?
In Fig.~\ref{UGJKsk}, top panel, we see the time evolution of $(u-g)_0$
and $(J110-K222)_0$ of
both the Salpeter (thick) - and the Kroupa (thin) models, including
absorption (solid) and without (dashed).
The bottom panel shows the differences in these colours between the
two models, both including absorption (solid) and without (dashed).
The thin dotted line shows zero level. \\
For the $(u-g)_0$ colour, we see the same conspiracy between SFR
and absorption as before. The colour difference in unabsorbed starlight
between the two models (bottom left panel dashed line), which was already
very small from the beginning (below 0.1 mag), is even diminished by the
absorption.
In $(J110-K222)_0$, on the other hand, the differences between the
Salpeter - and the Kroupa models increase when absorption is taken into
account, but
they remain too small to be measured (only up to 0.2 mags).
In colours made up of two magnitudes from widely separated wavelength
regions, such as $(u-K222)_0$, they even reach 0.5 mags, but in relation
to the larger variation of these colours, these differences are less
expressive than the 0.2 mags in $(J-K)_0$. \\
In Fig.~\ref{UGJKdfe}, we see the effect of reddening on the bluest
and the reddest colours of our study, $(u-g)_0$ and $(J110-K222)_0$,
respectively.
The top panels show the time evolution of these two colours for the
Kroupa model in all three viewing angles, as well as the absorptionless
case, which is the same for all viewing angles.
The bottom panel shows the differences between the absorbed and the
unabsorbed models in all three inclinations.
Clearly, the absorption effects on the bluest colours are
small, only up to $\sim 0.2$ mags in $(u-g)_0$.
In $(J110-K222)_0$, on the other hand, they reach up to
0.8 mags (edge-on),
so this colour might be suitable to detect orientation effects.
It may seem counter-intuitive, that inclination-induced reddening
on the bluest colour is much smaller than in the near infrared.
This must be because the differential extinction in $J110$ vs. $K222$
is much larger than in $u$ vs. $g$.

Having calculated galactic models both omitting dust effects and,
as a novelty, self-consistently including them, gives us the unique
occasion to test the usual assumption that the combined absorption and
reddening effects do not significantly alter colour-$M/L$ relations
(Bell \& de Jong \cite{bell}).
Fig.~\ref{colML} shows four such relations, combining colours and
$M/L$ ratios from the reddest and the bluest wavelength regions of
our calculations.
For the relations involving the $(u-g)_0$ colour (left panels),
the assumption works reasonably well. The dust effects move the colours
and the $M/L$ ratios along the main relations, thereby keeping them
in place.
For the relations involving $J110-K222$ (right panels), on the other
hand, the colour-$M/L$ relations are shifted by around 0.2 mags to
the red, when absorption is included (The dust affects this colour
much more than the $M/L$ ratios).
It seems difficult to keep these relations intact by altering the
colour and the $M/L$ ratios at the same time, since they are not 
linear in the first place.
A systematic investigation shows, that the assumption can be used
for colours involving passbands bluewards of the SDSS $i$ band.

\subsection{Apparent magnitudes}

In practice, apparent magnitudes (redshifted and corrected for distance)
are more relevant than intrinsic magnitudes, as they are the quantities
that are actually observed.
They are given in Tables~\ref{salpeterapparentsdss}
to~\ref{kroupaapparenthst} for the same bands as in
Tables~\ref{salpeterintrinsicsdss} to~\ref{kroupaintrinsichst}.
Due to the very low fluxes at the highest redshift (9.5116, corresponding
to a Universe age of 0.5 Gyr), the calculation of the magnitudes at
this age suffers too much from precision errors, so these values
should be taken with a grain of salt, especially in the bluer passbands.
In the bluest bands ($u$, $g$, $U336$, and $B439$), the errors might
even affect the second time step (redshift 5.6177) as well, especially
when combining the magnitudes to calculate colours, on which small
differences have a much more dramatic effect than on magnitudes. \\
For the $u$ and $K222$ bands, the redshift evolution is shown in
the top panels of Fig.~\ref{uksk}, again for both models in the diagonal
view, and both including - and not including absorption, like in the 
first and fourth panels of Fig.~\ref{MLUKsk}.
In the bottom panels, we see the differences between the Kroupa and the
Salpeter models.
These differences between the two models show the same tendencies
as for the intrinsic magnitudes ($\sim 0.5$ in the absorptionless case,
reduced when absorption is included), and are probably even harder to
detect than they would be in the intrinsic magnitudes, since they are
dominated by distance modulus effects.
Again, the $K222$ band is slightly better for detecting the
differences (they remain around 0.5 mag even after including
absorption) than the SLOAN filters, but still not good enough. \\
The absorption effects on the Kroupa model, that is the magnitude
differences between the absorptionless and the absorbed model,
are given in all calculated passbands in
Tables~\ref{ACkroupasdss} and~\ref{ACkroupahst} for the diagonal view.
Together with the inclination corrections given in
Tables~\ref{ICkroupasdss} to~\ref{ICekroupahst}, they can be calculated
for all three viewing angles.
By subtracting the inclination or absorption corrections for two
magnitudes, the inclination or absorption (reddening) corrections
for the corresponding colour can be derived.
This can be useful to make absorption corrections to galactic models.
We only give these corrections for the Kroupa model, because this model
is more realistic, and the corrections are very similar for the
Salpeter model.
As expected, the absorption is the strongest in the edge-on view and the
weakest face-on, but the latter does not differ much from the diagonal case.
So even for face-on galaxies, it will be impossible to infer the IMF from
an integrated magnitude, whereas for edge-on galaxies, the situation is
even worse. \\
If the apparent magnitudes calculated from the models cannot be used to
discriminate between the two different IMFs, do they at least reproduce
the empirical data?
In order to compare our model magnitudes with those from Section~\ref{sex}
that represent similar galaxies as in our models, we reduced both data sets
(edge-on and face-on) to those galaxies with similar sizes (Petrosian radii)
and structure (concentration indices) as the model galaxies.
Ideally, one should compare the models to galaxies with the same mass
and morphological type, but since these quantities are not given in the
SDSS, we resort to size and concentration parameters.
More precisely, we determined petroR90$_r$ and C$_r$ as a function of
redshift $z$ for both the Salpeter and the Kroupa models, face-on and edge-on,
and then reduced the data sets to those galaxies that fulfilled the following
criterion:
   \begin{displaymath}
      0.5\cdot{\rm petroR90}_{r,mod}(z) < {\rm petroR90}_{r,emp}(z)
   \end{displaymath}
   \begin{displaymath}
       < 2.0\cdot{\rm petroR90}_{r,mod}(z) \wedge
   \end{displaymath}
   \begin{equation}
   \label{subsample}
      0.98\cdot{\rm C}_{r,mod}(z) < {\rm C}_{r,emp}(z) <1.02\cdot{\rm C}_{r,mod}(z)
   \end{equation}
where petroR90$_{r,mod}(z)$ and C$_{r,mod}(z)$ are the Petrosian radius and
concentration index calculated from the models and petroR90$_{r,emp}(z)$ and
C$_{r,emp}(z)$ are the values taken from the SDSS data base, as well as the
redshift $z$.
The calculated Petrosian radii (both in kpc and in arcsec) and
concentration indices are given for the Kroupa model in
Tables~\ref{R90kroupa} to~\ref{Ckroupa}, as they could be useful to
identify Milky Way-type galaxies or progenitors at high redshift.
At the first time step (0.5 Gyr, redshift 9.5116), however, these
values still suffer from the initial border conditions of the model,
and cannot be used.
The concentration indices could only be calculated from redshift
1.5915 on. \\
The apparent $u$ band magnitudes including their error bars of the
subsamples are plotted as a function of redshift in Fig.~\ref{uskS}.
The thick dashed and dotted lines show the evolution of the corresponding
models (dashed: face-on, dotted: edge-on). The thin lines show the respective
other views for the same models (thus the thick lines in the upper panels
correspond to the thin lines in the lower panels, and vice-versa), to show
the differences in brightness between the two viewing angles.
The models are only shown until the second-last time step at
redshift 0.0369, as the magnitudes at redshift 0 depend on the actual
distance, which is unknown.
The distance modulus $m-M$ of 25 used in Tables~\ref{salpeterapparentsdss}
to~\ref{kroupaapparenthst} is only a convention, whereas in reality the
distance modulus at zero redshift depends upon the actual distance and
equation~\ref{mM} is not valid. \\
As can be seen from Fig.~\ref{uskS}, the models represent the $u$ magnitudes
of the empirical samples well, and they even reproduce the brightness
differences between the face-on and the edge-on views. \\
Fig.~\ref{rskS} shows the same galaxies and models as Fig.~\ref{uskS}, but
in the $r$ band. Here too, the agreement is good, although one could
argue that the magnitudes of the Salpeter model are a bit too faint.

\subsection{Apparent colours}

\begin{figure*}
\centering
\includegraphics{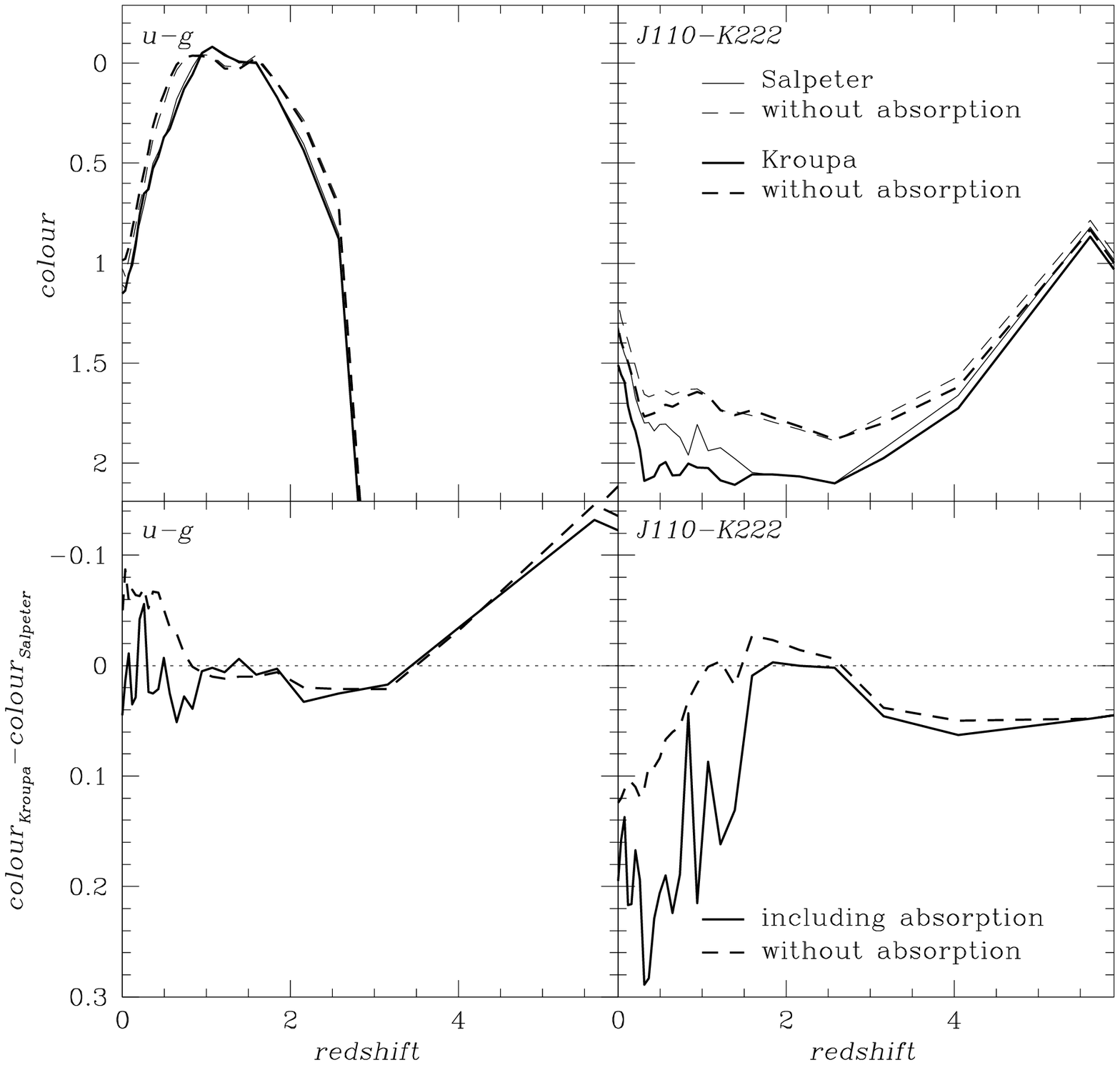}
\caption{Top panels: Apparent $u-g$ and $J110-K222$ colour evolution of the
       Salpeter (thin) and the Kroupa (thick) models as a function of redshift,
       both including (solid) and omitting (dashed) gas absorption.
       Bottom panels: differences in the $u-g$ and $J110-K222$ colours between
       the Salpeter and the Kroupa models, both including (solid) and omitting
       (dashed) gas absorption.
       The thin dotted lines show the zero level.}
\label{ugjksk}
\end{figure*}

\begin{figure*}
\centering
\includegraphics{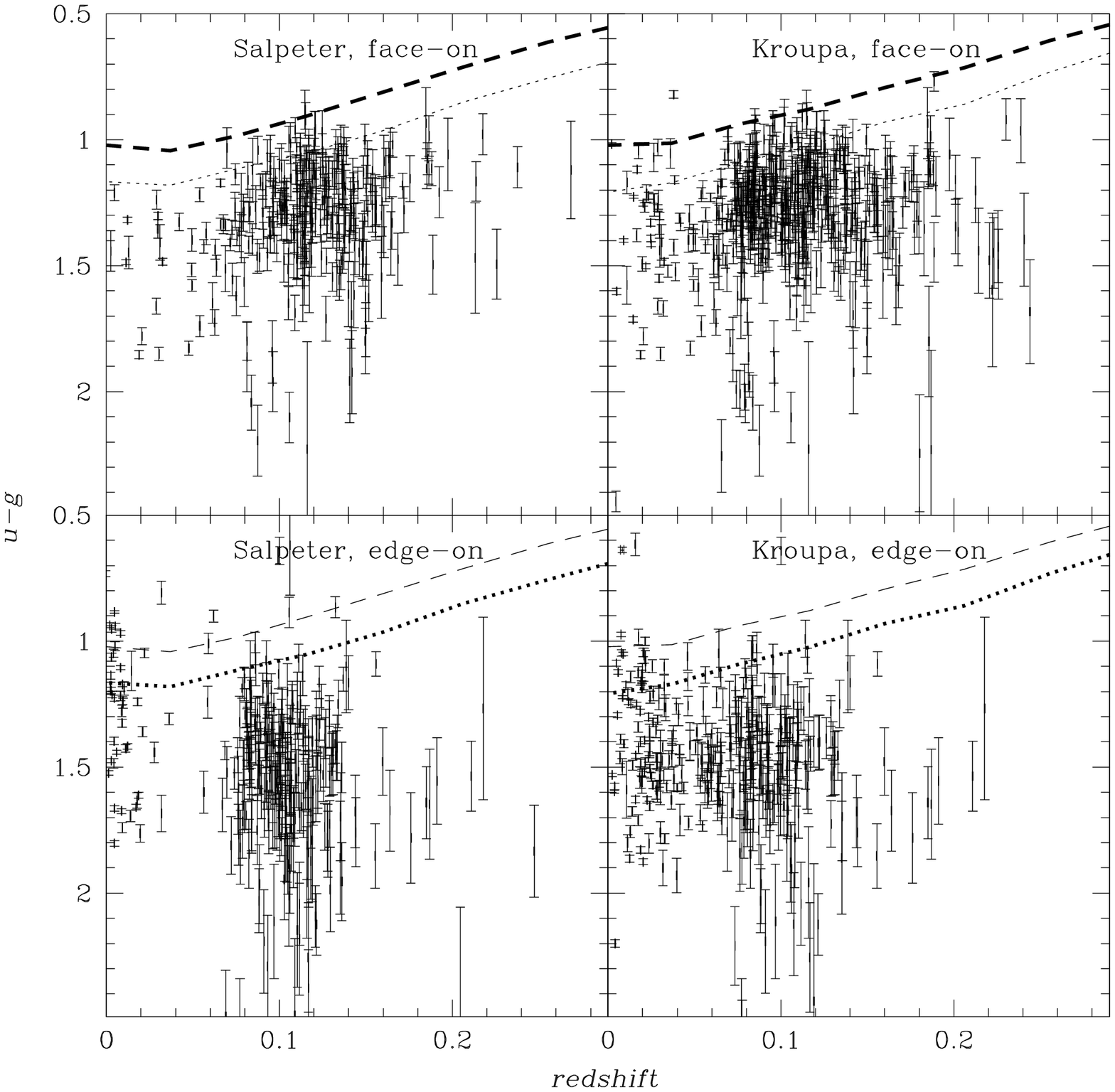}
\caption{Apparent $u-g$ colours of the Salpeter and Kroupa models as seen
       face-on (dashed) and edge-on (dotted) as a function of redshift.
       Overlaid are the SDSS DR4 data with their observational errors of
       galaxies that have similar sizes, concentration parameters, and
       orientation as the models (for a more precise description, see the
       text).}
\label{ugskS}
\end{figure*}

\begin{figure*}
\centering
\includegraphics{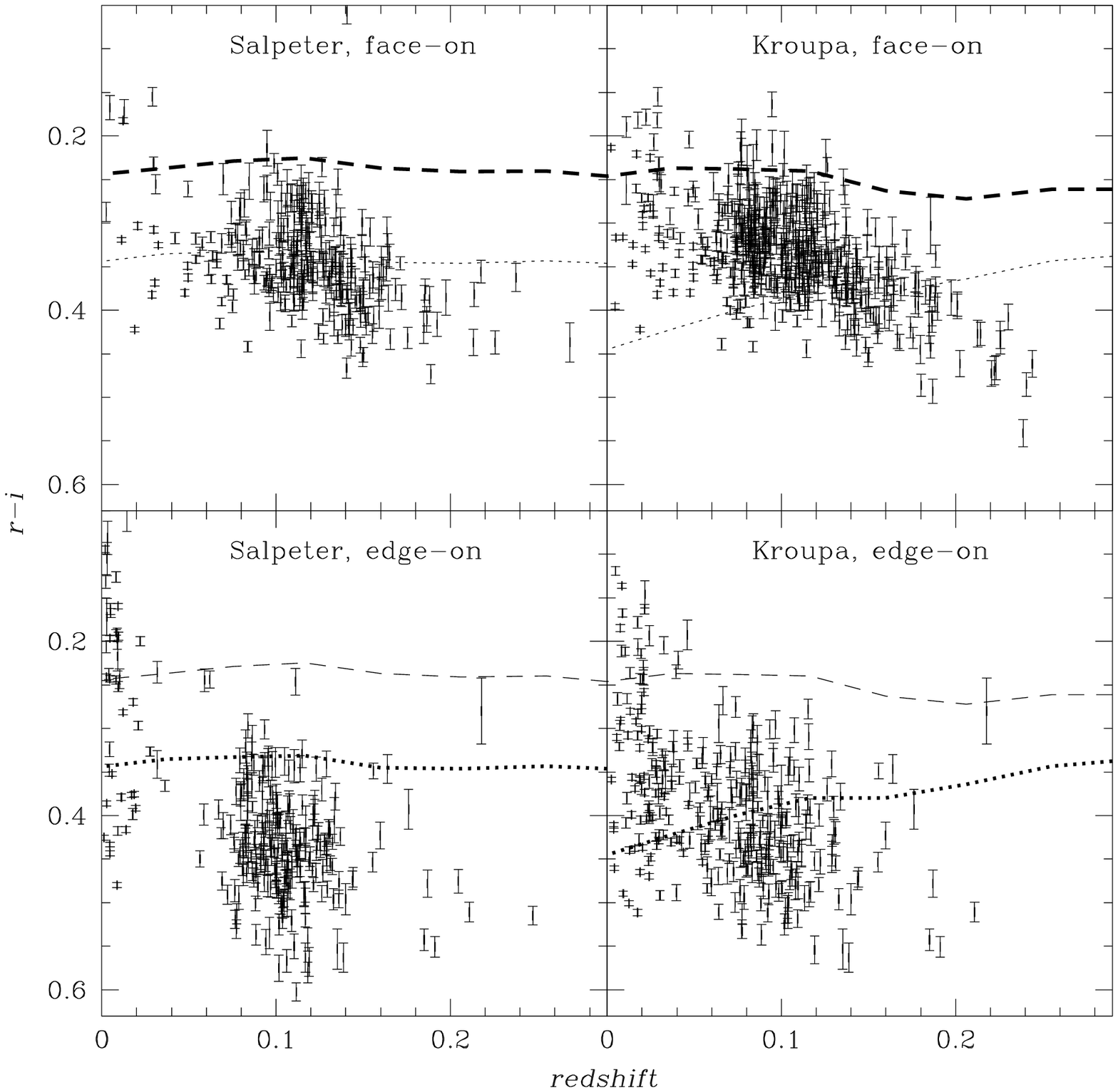}
\caption{Like Fig.~\ref{ugskS}, but for $r-i$.}
    \label{riskS}
\end{figure*}

\begin{figure*}
\centering
\includegraphics{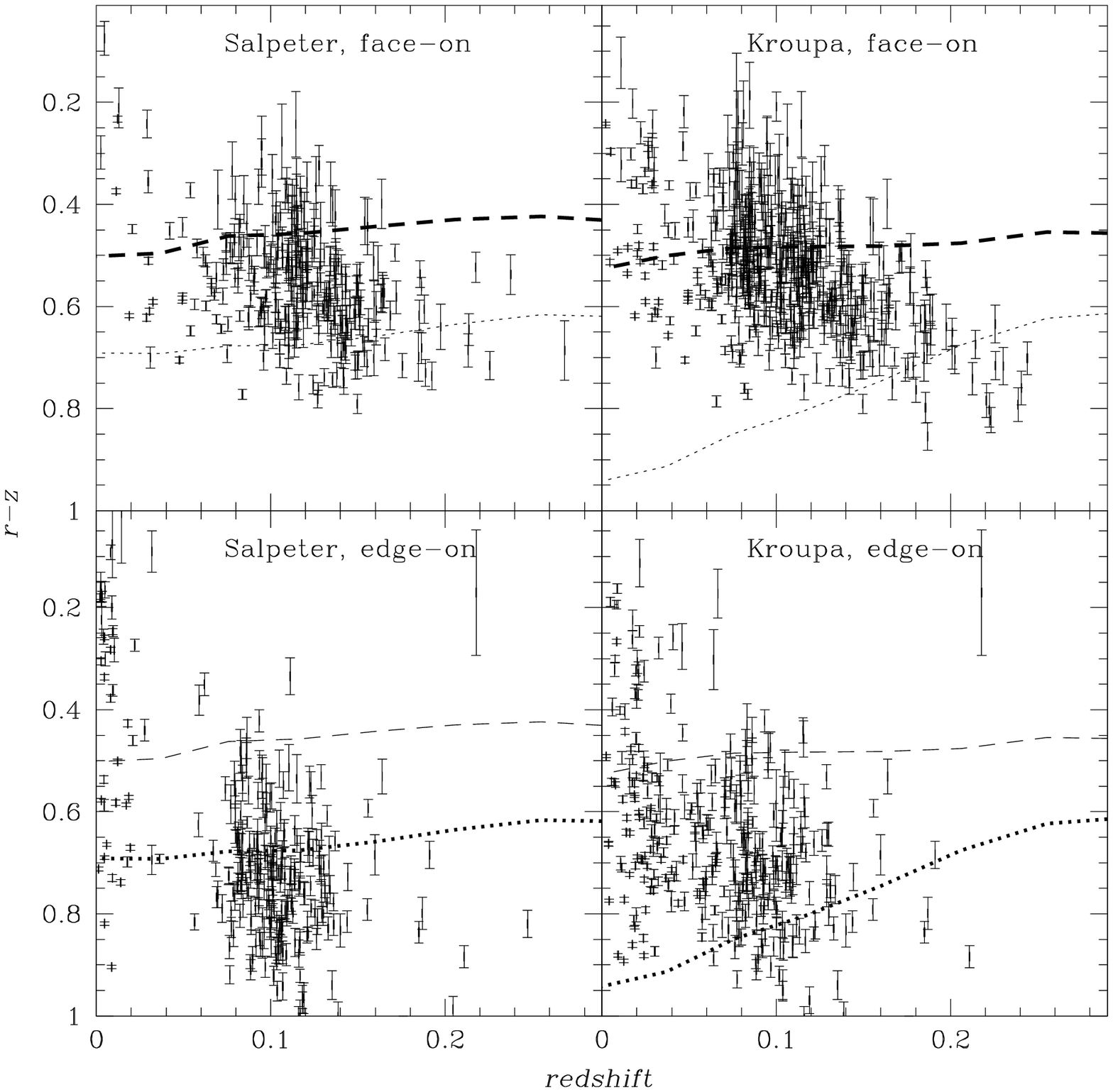}
\caption{Like Fig.~\ref{ugskS}, but for $r-z$.}
    \label{rzskS}
\end{figure*}

Let us move on to apparent colours.
Fig.~\ref{ugjksk} shows the redshift evolution of apparent $u-g$ and
$J110-K222$, the different line types and widths having the same
meaning as in Fig.~\ref{uksk}.
The figure is only shown out to redshift 5.9, due to the
above-mentioned imprecisions at redshift 9.5116. In $u-g$, they might
even affect the data point at redshift 5.6177.
In $u-g$, we again see our ``IMF degeneracy'', the fact that the higher
absorption in the Kroupa model compensates the bluer colour caused by the
higher SFR.
This leaves the $u-g$ differences between the two models at a level
of at most 0.05 mags, which is undetectable given the much larger
variations this colour shows during the evolution.
At best, the differences could be seen in $J110-K222$, where they
amount to $\sim 0.2$ mags (edge-on up to 0.4 mags), which is a significant
fraction of the variation of this colour after 2.5 Gyr. ($\sim 0.7$ mags).
Let us see how the models compare to the empirical data.
In Figs.~\ref{ugskS} to~\ref{rzskS}, we see the $u-g$, $r-i$, and $r-z$
colours of the two models, face-on and edge-on, as a function of redshift
in the same coding as in Fig.~\ref{uskS}.
Overplotted are the colours including error bars of the four subsamples
of galaxies selected using equation~\ref{subsample}, so the models are only
compared to data from galaxies of the same types (angular sizes,
concentration parameters, and viewing angles). \\
Obviously, the model colours are too blue by about 0.3 to 0.4
mags in $u-g$, by about 0.1 mags in $r-i$, and more or less
in the right colour range for $r-z$.
Also, the trends with redshift are not always well reproduced in the
latter two colours.
This could be an indication, that hierarchical-based accretion
histories may induce a delayed galaxy formation, resulting in too blue
colours especially at high redshift.
It looks like SDSS galaxies on average have an earlier Hubble type
star formation history than the Milky Way type model galaxy.
This argument is also supported by the fact, that the monolithic
collapse model in Westera et al. \cite{westera} produces colours in better
agreement with the data.
In spite of these systematic deviations, the models do seem to reproduce
relative properties, i. e. the effect of galaxy orientation on integrated
colours, as can be seen from the fact, that they nicely reproduce
the shift between face-on - and edge-on colours, thereby underlining
the usefulness of the corrections given in Tables~\ref{ACkroupasdss}
to~\ref{ICekroupahst}.
This is why we believe that the results we found comparing the models with
different IMFs are also realistic.
We thus conclude that it is difficult to make statements about the stellar
IMF from integrated colours or magnitudes of galaxies.
Our best bet for this purpose are infrared colours, i. e. $J110-K222$.

\section{Summary and Conclusions}
\label{summary}

In this work, we use chemo-dynamical models to investigate the influence of
the Initial Mass Function (IMF) on the evolution of a Milky Way-type
disk galaxy, in particular of its colours. \\
For this purpose, we developed two chemodynamical models of such a galaxy
with the same boundary conditions, but using different IMFs: Salpeter and
Kroupa, which differ in their low-to-high mass stars ratios.
The Kroupa model, having a higher fraction of high mass stars, begins
with a lower SFR than the Salpeter model, but from 5 Gyr on, this reverses.
The Kroupa model also has a higher gas density and metallicity than the
Salpeter model at all ages. \\
With these two models, we performed a spectral analysis, evaluated
with a state of the art evolutionary code and spectral library. The
programme transforming the models into spectral properties takes
into account the three-dimensional distribution of the stars and
the interstellar matter. It includes internal gas absorption and
re-emission and is also able to include foreground reddening.
We obtain two-dimensional HST (WFPC2 and NICMOS), and SDSS $ugriz$
images of the model galaxies, giving intrinsic and apparent
magnitudes and colours in up to $320 \times 320$ pixels.
We also obtain intrinsic and apparent integrated spectra and
colours of the model galaxies.
All of these quantities were calculated with a time resolution of
$0.5$~Gyr.
The programme is able to view the model galaxies from different angles
(diagonally, face-on, and edge-on), which allows to analyse orientation
effects on the spectral properties.
Furthermore, by recalculating the models artificially omitting the
gas absorption, we could disentangle absorption effects from other effects.
We provide photometric absorption and inclination corrections
in the SDSS $ugriz$ and the HST WFPC2 and NICMOS systems
(Tables~\ref{ACkroupasdss} to~\ref{ICekroupahst}).

We find, that the effect of the IMF on the internal gas absorption
is larger than its effect on the light from the stellar content.
However, the two effects work in the opposite sense (An IMF with more
high mass stars leads to brighter and bluer stellar light, but also to
more interstellar dust and thus to more absorption), causing a kind
of ``IMF degeneracy''.
The most likely wavelength region, in which to detect IMF effects
is the infrared (i. e. $JHK$).
Here, the differences between the two models amount to $\sim 0.5$ mags
in $K222$, and to $\sim 0.2$ mags in $J110-K222$. \\
The effects of inclination on a galaxy's magnitudes and colours,
on the other hand, are larger than the ones due to the IMF.
Seeing a galaxy edge-on instead of face-on or diagonally can make it appear
up to 1.5 mags fainter in $u$, or 1 mag in $z$, or 0.6 mags redder in
$J110-K222$. \\
A comparison of the calculated integrated model magnitudes and colours
with SDSS data partly shows good agreement, especially in the
$u$ and $g$ magnitudes, and the $riz$ colours.
There are some systematic deviations in the UV colours, indicating
that SDSS galaxies might have on average an earlier Hubble type
star formation history than the model galaxy, which is more like
the Milky Way.
On the other hand, the relative tendencies in these
colours are well reproduced (i. e. the shift between face-on and edge-on
colours of galaxies of the same type).
We conclude from this, that the theoretical results presented in the two
previous paragraphs should hold true in practice as well.

As a side study, we verify the assumption by Bell \& de Jong
\cite{bell}, that the combined absorption and reddening effects of
dust do not significantly alter the colour-$M/L$ relations for galaxy
models, and find this assumption only to be true for colours
involving passbands in the bluer wavelength ranges
(bluewards of the $I$ band).

\begin{acknowledgements}
   This work was supported by the Swiss National Science Foundation.
\end{acknowledgements}

\appendix

\section{Intrinsic magnitudes}

   \begin{table*}
    \begin{center}
     \caption[Intrinsic integrated SDSS magnitudes of the Salpeter model]
    {Intrinsic integrated SDSS magnitudes of the Salpeter model.}
     \label{salpeterintrinsicsdss}
    $$ 
\begin{array}{rrrrrrrr}
      \hline
 t [Gyr] & M_{Bol} & M_{V_j} &   M_{u} &   M_{g} &   M_{r} &   M_{i} &   M_{z} \\
            \hline
     0.5 & -15.586 & -12.621 & -13.207 & -12.918 & -12.430 & -12.171 & -11.957 \\
     1.0 & -19.467 & -17.749 & -17.563 & -17.844 & -17.706 & -17.643 & -17.583 \\
     1.5 & -20.513 & -19.160 & -18.736 & -19.187 & -19.157 & -19.147 & -19.141 \\
     2.0 & -21.132 & -19.964 & -19.403 & -19.944 & -19.986 & -20.013 & -20.047 \\
     2.5 & -21.562 & -20.505 & -19.843 & -20.449 & -20.548 & -20.609 & -20.682 \\
     3.0 & -21.848 & -20.855 & -20.114 & -20.770 & -20.916 & -21.008 & -21.117 \\
     3.5 & -22.004 & -21.067 & -20.242 & -20.954 & -21.145 & -21.261 & -21.397 \\
\hline
         \end{array}
     $$ 
   \end{center}
   \end{table*}

   \begin{table*}
    \begin{center}
     \caption[Intrinsic integrated HST magnitudes of the Salpeter model]
    {Intrinsic integrated HST magnitudes of the Salpeter model.}
     \label{salpeterintrinsichst}
    $$ 
     \begin{array}{rrrrrrrrrr}
      \hline
 t [Gyr] & M_{U336} & M_{B439} & M_{V555} & M_{V606} & M_{R675} & M_{I814} & M_{J110} & M_{H160} & M_{K222} \\
            \hline
         0.5 & -14.51 & -12.94 & -12.70 & -12.65 & -12.59 & -12.55 & -12.48 & -12.49 & -12.52 \\
         1.0 & -18.73 & -17.77 & -17.77 & -17.83 & -17.96 & -18.10 & -18.20 & -18.58 & -18.75 \\
         1.5 & -19.86 & -19.08 & -19.16 & -19.26 & -19.43 & -19.64 & -19.80 & -20.29 & -20.50 \\
         2.0 & -20.51 & -19.81 & -19.95 & -20.07 & -20.28 & -20.52 & -20.73 & -21.30 & -21.54 \\
         2.5 & -20.94 & -20.30 & -20.49 & -20.63 & -20.85 & -21.14 & -21.40 & -22.03 & -22.30 \\
         3.0 & -21.20 & -20.60 & -20.83 & -20.99 & -21.23 & -21.56 & -21.86 & -22.57 & -22.87 \\
         3.5 & -21.31 & -20.77 & -21.03 & -21.21 & -21.47 & -21.83 & -22.17 & -22.93 & -23.26 \\
\hline
         \end{array}
     $$ 
   \end{center}
   \end{table*}

   \begin{table*}
    \begin{center}
     \caption[Intrinsic integrated SDSS magnitudes of the Kroupa model]
    {Intrinsic integrated SDSS magnitudes of the Kroupa model.}
     \label{kroupaintrinsicsdss}
    $$ 
     \begin{array}{rrrrrrrr}
      \hline
 t [Gyr] & M_{Bol} & M_{V_j} &   M_{u} &   M_{g} &   M_{r} &   M_{i} &   M_{z} \\
            \hline
     0.5 & -15.823 & -13.180 & -13.541 & -13.414 & -12.994 & -12.714 & -12.490 \\
     1.0 & -19.526 & -17.907 & -17.659 & -17.993 & -17.839 & -17.751 & -17.705 \\
     1.5 & -20.522 & -19.227 & -18.769 & -19.252 & -19.203 & -19.179 & -19.189 \\
     2.0 & -21.098 & -19.968 & -19.380 & -19.950 & -19.975 & -19.996 & -20.047 \\
     2.5 & -21.482 & -20.434 & -19.757 & -20.380 & -20.466 & -20.528 & -20.618 \\
     3.0 & -21.762 & -20.777 & -20.025 & -20.695 & -20.830 & -20.924 & -21.046 \\
     3.5 & -21.937 & -20.966 & -20.147 & -20.859 & -21.039 & -21.161 & -21.315 \\
\hline
         \end{array}
     $$ 
   \end{center}
   \end{table*}

   \begin{table*}
    \begin{center}
     \caption[Intrinsic integrated HST magnitudes of the Kroupa model]
    {Intrinsic integrated HST magnitudes of the Kroupa model.}
     \label{kroupaintrinsichst}
    $$ 
     \begin{array}{rrrrrrrrrr}
      \hline
 t [Gyr] & M_{U336} & M_{B439} & M_{V555} & M_{V606} & M_{R675} & M_{I814} & M_{J110} & M_{H160} & M_{K222} \\
            \hline
         0.5 & -14.80 & -13.39 & -13.24 & -13.20 & -13.14 & -13.09 & -13.02 & -13.04 & -13.07 \\
         1.0 & -18.82 & -17.91 & -17.93 & -17.98 & -18.07 & -18.22 & -18.35 & -18.78 & -18.95 \\
         1.5 & -19.89 & -19.14 & -19.23 & -19.32 & -19.47 & -19.67 & -19.87 & -20.41 & -20.63 \\
         2.0 & -20.49 & -19.81 & -19.96 & -20.07 & -20.26 & -20.51 & -20.76 & -21.37 & -21.63 \\
         2.5 & -20.85 & -20.23 & -20.42 & -20.55 & -20.77 & -21.07 & -21.36 & -22.05 & -22.33 \\
         3.0 & -21.11 & -20.53 & -20.75 & -20.90 & -21.15 & -21.48 & -21.82 & -22.57 & -22.89 \\
         3.5 & -21.23 & -20.68 & -20.93 & -21.10 & -21.37 & -21.74 & -22.11 & -22.94 & -23.30 \\
\hline
         \end{array}
     $$ 
   \end{center}
   \end{table*}

   \begin{table*}
    \begin{center}
     \caption[Apparent integrated SDSS magnitudes of the Salpeter model]
    {Apparent integrated SDSS magnitudes of the Salpeter model.}
     \label{salpeterapparentsdss}
    $$ 
     \begin{array}{rrrrrrrrrr}
      \hline
 t [Gyr] &    redshift &       m-M & m_{Bol} & m_{V_j} &   m_{u} &   m_{g} &   m_{r} &   m_{i} &   m_{z} \\
            \hline
   0.5 & 9.5116 & 49.953869 &  39.968 & 161.915 & 162.836 & 162.254 & 161.664 & 161.259 &  52.997 \\
   1.0 & 5.6177 & 48.640133 &  33.535 &  39.367 &  59.169 &  56.694 &  36.545 &  34.365 &  32.700 \\
   1.5 & 4.0444 & 47.801151 &  30.919 &  31.566 &  48.184 &  32.844 &  31.073 &  30.684 &  30.748 \\
   2.0 & 3.1576 & 47.159698 &  29.177 &  29.381 &  34.035 &  29.857 &  29.311 &  29.372 &  29.479 \\
   2.5 & 2.5758 & 46.626331 &  27.854 &  28.294 &  29.258 &  28.404 &  28.318 &  28.430 &  28.454 \\
   3.0 & 2.1589 & 46.160126 &  26.821 &  27.528 &  27.922 &  27.517 &  27.579 &  27.665 &  27.552 \\
   3.5 & 1.8422 & 45.738579 &  26.005 &  26.979 &  27.111 &  26.944 &  27.052 &  26.996 &  26.843 \\
\hline
         \end{array}
     $$ 
   \end{center}
   \end{table*}

   \begin{table*}
    \begin{center}
     \caption[Apparent integrated HST magnitudes of the Salpeter model]
    {Apparent integrated HST magnitudes of the Salpeter model.}
     \label{salpeterapparenthst}
    $$ 
     \begin{array}{rrrrrrrrrr}
      \hline
 redshift & m_{U336} & m_{B439} & m_{V555} & m_{V606} & m_{R675} & m_{I814} & m_{J110} & m_{H160} & m_{K222} \\
            \hline
 9.5116 & 161.8 & 162.6 &   162 & 161.7 & 161.2 & 58.57 & 39.79 & 36.96 & 36.71 \\
 5.6177 & 57.42 & 59.80 & 39.05 & 36.33 & 34.78 & 32.76 & 32.09 & 31.68 & 31.27 \\
 4.0444 & 47.64 & 36.83 & 31.89 & 31.16 & 30.42 & 30.23 & 30.11 & 29.59 & 28.45 \\
 3.1576 & 39.10 & 30.20 & 29.54 & 29.30 & 29.06 & 28.94 & 28.76 & 27.76 & 26.83 \\
 2.5758 & 28.93 & 28.73 & 28.30 & 28.20 & 28.08 & 27.96 & 27.63 & 26.27 & 25.53 \\
 2.1589 & 26.85 & 27.62 & 27.53 & 27.46 & 27.36 & 27.11 & 26.57 & 25.28 & 24.50 \\
 1.8422 & 26.07 & 27.07 & 26.98 & 26.90 & 26.78 & 26.42 & 25.70 & 24.48 & 23.64 \\
\hline
         \end{array}
     $$ 
   \end{center}
   \end{table*}

   \begin{table*}
    \begin{center}
     \caption[Apparent integrated SDSS magnitudes of the Kroupa model]
    {Apparent integrated SDSS magnitudes of the Kroupa model.}
     \label{kroupaapparentsdss}
    $$ 
     \begin{array}{rrrrrrrr}
      \hline
 t [Gyr] & m_{Bol} & m_{V_j} & m_{u} &   m_{g} &   m_{r} &   m_{i} &   m_{z} \\
            \hline
   0.5 &  39.713 & 161.914 & 162.834 & 162.252 & 161.663 & 161.257 &  52.725 \\
   1.0 &  33.463 &  39.350 &  59.066 &  56.723 &  36.533 &  34.342 &  32.664 \\
   1.5 &  30.902 &  31.608 &  48.202 &  32.899 &  31.100 &  30.697 &  30.752 \\
   2.0 &  29.206 &  29.453 &  34.143 &  29.948 &  29.375 &  29.425 &  29.524 \\
   2.5 &  27.931 &  28.400 &  29.396 &  28.517 &  28.417 &  28.521 &  28.542 \\
   3.0 &  26.905 &  27.635 &  28.073 &  27.635 &  27.681 &  27.761 &  27.638 \\
   3.5 &  26.070 &  27.051 &  27.189 &  27.019 &  27.122 &  27.068 &  26.925 \\
\hline
         \end{array}
     $$ 
   \end{center}
   \end{table*}

   \begin{table*}
    \begin{center}
     \caption[Apparent integrated HST magnitudes of the Kroupa model]
    {Apparent integrated HST magnitudes of the Kroupa model.}
     \label{kroupaapparenthst}
    $$ 
     \begin{array}{rrrrrrrrrr}
      \hline
 redshift & m_{U336} & m_{B439} & m_{V555} & m_{V606} & m_{R675} & m_{I814} & m_{J110} & m_{H160} & m_{K222} \\
            \hline
 9.5116 & 161.8 & 162.6 &   162 & 161.7 & 161.2 & 58.14 & 39.58 & 36.74 & 36.49 \\
 5.6177 & 57.31 & 59.76 & 39.03 & 36.31 & 34.76 & 32.72 & 32.05 & 31.63 & 31.19 \\
 4.0444 & 47.65 & 36.89 & 31.93 & 31.19 & 30.44 & 30.24 & 30.11 & 29.57 & 28.39 \\
 3.1576 & 39.21 & 30.30 & 29.62 & 29.37 & 29.12 & 28.99 & 28.80 & 27.78 & 26.83 \\
 2.5758 & 29.07 & 28.85 & 28.40 & 28.31 & 28.17 & 28.05 & 27.72 & 26.34 & 25.62 \\
 2.1589 & 27.00 & 27.75 & 27.65 & 27.57 & 27.46 & 27.21 & 26.65 & 25.36 & 24.59 \\
 1.8422 & 26.15 & 27.15 & 27.05 & 26.97 & 26.85 & 26.50 & 25.79 & 24.58 & 23.74 \\
\hline
         \end{array}
     $$ 
   \end{center}
   \end{table*}

\section{Mass-to-light ratios}

   \begin{table*}
    \begin{center}
     \caption[Stellar mass-to-light ratios of the Salpeter model]
    {Stellar mass-to-light ratios of the Salpeter model.}
     \label{salpeterML}
    $$ 
     \begin{array}{rrrrrrrrrrr}
      \hline
 t [Gyr] & M/L_{Bol} & M/L_{V_j} & M/L_{u} & M/L_{g} & M/L_{r} & M/L_{i} & M/L_{z} & M/L_{J110} & M/L_{H160} & M/L_{K222} \\
            \hline
  0.5 & 0.0075 & 0.1145 & 0.0137 & 0.0667 & 0.1568 & 0.2202 & 0.2757 & 0.4439 & 0.2919 & 0.5347 \\
  1.0 & 0.0367 & 0.1787 & 0.0435 & 0.1253 & 0.2134 & 0.2503 & 0.2719 & 0.4022 & 0.1886 & 0.3014 \\
  1.5 & 0.0720 & 0.2502 & 0.0759 & 0.1869 & 0.2881 & 0.3218 & 0.3326 & 0.4742 & 0.2012 &  0.309 \\
  2.0 & 0.1098 &  0.322 & 0.1107 & 0.2511 & 0.3623 &  0.391 & 0.3896 & 0.5398 & 0.2143 & 0.3208 \\
  2.5 & 0.1534 & 0.4061 & 0.1533 & 0.3274 & 0.4482 & 0.4689 & 0.4507 & 0.6079 & 0.2261 & 0.3298 \\
  3.0 & 0.2108 & 0.5262 & 0.2136 & 0.4357 & 0.5711 & 0.5807 & 0.5399 & 0.7079 & 0.2463 &  0.348 \\
  3.5 & 0.2838 & 0.6727 &  0.295 & 0.5715 & 0.7188 & 0.7148 & 0.6483 & 0.8321 & 0.2748 & 0.3793 \\
\hline
         \end{array}
     $$ 
   \end{center}
   \end{table*}

   \begin{table*}
    \begin{center}
     \caption[Stellar mass-to-light ratios of the Kroupa model]
    {Stellar mass-to-light ratios of the Kroupa model.}
     \label{kroupaML}
    $$ 
     \begin{array}{rrrrrrrrrrr}
      \hline
 t [Gyr] & M/L_{Bol} & M/L_{V_j} & M/L_{u} & M/L_{g} & M/L_{r} & M/L_{i} & M/L_{z} & M/L_{J110} & M/L_{H160} & M/L_{K222} \\
            \hline
  0.5 & 0.0056 & 0.0643 & 0.0095 & 0.0397 & 0.0876 & 0.1255 & 0.1586 & 0.2544 & 0.1664 & 0.3011 \\
  1.0 & 0.0288 & 0.1279 & 0.0330 & 0.0905 & 0.1563 & 0.1876 & 0.2012 & 0.2895 &   0.13 & 0.2078 \\
  1.5 & 0.0568 & 0.1873 & 0.0586 & 0.1401 & 0.2198 & 0.2487 & 0.2533 & 0.3516 & 0.1428 & 0.2188 \\
  2.0 & 0.0878 & 0.2487 & 0.0877 & 0.1936 & 0.2837 & 0.3079 &  0.302 & 0.4078 & 0.1545 & 0.2295 \\
  2.5 & 0.1242 & 0.3261 & 0.1248 & 0.2624 & 0.3635 &   0.38 & 0.3595 & 0.4722 & 0.1671 & 0.2406 \\
  3.0 &  0.168 & 0.4162 & 0.1706 & 0.3436 & 0.4551 & 0.4618 & 0.4243 & 0.5431 & 0.1808 & 0.2524 \\
  3.5 &  0.225 & 0.5502 &   0.24 & 0.4649 & 0.5907 & 0.5842 & 0.5211 & 0.6501 & 0.2029 & 0.2727 \\
\hline
         \end{array}
     $$ 
   \end{center}
   \end{table*}

   \begin{table*}
    \begin{center}
     \caption[Stellar mass-to-light ratios of the absorptionless Salpeter model]
    {Stellar mass-to-light ratios of the absorptionless Salpeter model.}
     \label{salpeterMLwoabs}
    $$ 
     \begin{array}{rrrrrrrrrrr}
      \hline
 t [Gyr] & M/L_{Bol} & M/L_{V_j} & M/L_{u} & M/L_{g} & M/L_{r} & M/L_{i} & M/L_{z} & M/L_{J110} & M/L_{H160} & M/L_{K222} \\
            \hline
  0.5 & 0.0075 & 0.1145 & 0.0137 & 0.0667 & 0.1568 & 0.2202 & 0.2757 & 0.4439 & 0.2919 & 0.5347 \\
  1.0 & 0.0339 & 0.1725 & 0.0413 & 0.1201 &  0.207 & 0.2443 & 0.2672 & 0.3966 & 0.1874 & 0.3006 \\
  1.5 & 0.0594 & 0.2257 & 0.0658 & 0.1655 & 0.2633 & 0.2998 & 0.3153 & 0.4541 & 0.1975 & 0.3062 \\
  2.0 & 0.0810 & 0.2681 & 0.0862 & 0.2024 & 0.3081 & 0.3434 &  0.353 & 0.4991 & 0.2071 & 0.3158 \\
  2.5 & 0.0995 & 0.3038 & 0.1041 & 0.2343 &  0.346 & 0.3801 & 0.3839 & 0.5349 & 0.2133 & 0.3208 \\
  3.0 & 0.1213 & 0.3461 &  0.125 & 0.2714 & 0.3908 & 0.4238 & 0.4214 & 0.5796 & 0.2238 & 0.3323 \\
  3.5 & 0.1557 & 0.4113 &  0.159 & 0.3292 & 0.4586 & 0.4896 & 0.4788 & 0.6501 & 0.2433 & 0.3573 \\
\hline
         \end{array}
     $$ 
   \end{center}
   \end{table*}

   \begin{table*}
    \begin{center}
     \caption[Stellar mass-to-light ratios of the absorptionless Kroupa model]
    {Stellar mass-to-light ratios of the absorptionless Kroupa model.}
     \label{kroupaMLwoabs}
    $$ 
     \begin{array}{rrrrrrrrrrr}
      \hline
 t [Gyr] & M/L_{Bol} & M/L_{V_j} & M/L_{u} & M/L_{g} & M/L_{r} & M/L_{i} & M/L_{z} & M/L_{J110} & M/L_{H160} & M/L_{K222} \\
            \hline
  0.5 & 0.0056 & 0.0643 & 0.0095 & 0.0397 & 0.0876 & 0.1255 & 0.1586 & 0.2544 & 0.1664 & 0.3011 \\
  1.0 & 0.0264 & 0.1229 & 0.0311 & 0.0863 &  0.151 & 0.1825 & 0.1972 &  0.285 & 0.1291 & 0.2071 \\
  1.5 & 0.0461 &  0.167 & 0.0500 & 0.1225 & 0.1988 & 0.2295 & 0.2386 & 0.3352 & 0.1398 & 0.2166 \\
  2.0 & 0.0638 & 0.2034 & 0.0668 & 0.1532 & 0.2373 & 0.2665 & 0.2707 &  0.374 & 0.1485 & 0.2253 \\
  2.5 & 0.0782 & 0.2356 & 0.0814 & 0.1807 &  0.271 & 0.2985 & 0.2982 & 0.4068 & 0.1558 & 0.2326 \\
  3.0 & 0.0934 & 0.2677 & 0.0965 & 0.2088 & 0.3043 &   0.33 & 0.3252 & 0.4382 &  0.163 & 0.2397 \\
  3.5 &  0.114 & 0.3103 & 0.1176 &  0.246 & 0.3478 &  0.371 & 0.3605 & 0.4801 & 0.1737 &  0.252 \\
\hline
         \end{array}
     $$ 
   \end{center}
   \end{table*}

\section{Correction tables}

   \begin{table*}
    \begin{center}
     \caption[Absorption corrections for the Kroupa model in the SDSS system]
     {Absorption corrections $m_{X,including abs.}-m_{X,abs.less}$ for the Kroupa model in the SDSS system.}
     \label{ACkroupasdss}
    $$ 
     \begin{array}{rrrrrrrr}
      \hline
 redshift & BOL & V_j & u & g & r & i & z \\
            \hline
 9.5116 &     0 &     0 &     0 &     0 &     0 &     0 &     0 \\
 5.6177 & 0.081 & 0.152 & 0.258 & 0.151 & 0.151 & 0.124 & 0.105 \\
 4.0444 & 0.210 & 0.339 & 0.498 & 0.378 & 0.294 & 0.259 & 0.254 \\
 3.1576 & 0.331 & 0.447 & 0.631 & 0.506 & 0.420 & 0.412 & 0.458 \\
 2.5758 & 0.488 & 0.627 & 0.827 & 0.669 & 0.620 & 0.677 & 0.607 \\
 2.1589 & 0.625 & 0.820 & 0.974 & 0.840 & 0.848 & 0.849 & 0.711 \\
 1.8422 & 0.727 & 0.992 & 1.037 & 0.966 & 1.045 & 0.912 & 0.824 \\
 1.5915 & 0.783 & 1.157 & 1.123 & 1.094 & 1.117 & 0.959 & 0.876 \\
 1.3868 & 0.828 & 1.247 & 1.187 & 1.224 & 1.121 & 0.998 & 0.911 \\
 1.2156 & 0.820 & 1.163 & 1.166 & 1.233 & 1.060 & 0.965 & 0.893 \\
 1.0696 & 0.837 & 1.105 & 1.179 & 1.234 & 1.027 & 0.934 & 0.878 \\
 0.9432 & 0.886 & 1.147 & 1.278 & 1.291 & 1.081 & 0.980 & 0.917 \\
 0.8322 & 0.757 & 0.991 & 1.241 & 1.147 & 0.918 & 0.818 & 0.736 \\
 0.7336 & 0.880 & 1.094 & 1.339 & 1.183 & 1.044 & 0.999 & 0.938 \\
 0.6453 & 0.908 & 1.101 & 1.402 & 1.178 & 1.052 & 1.016 & 0.941 \\
 0.5655 & 0.962 & 1.115 & 1.446 & 1.195 & 1.070 & 1.022 & 0.945 \\
 0.4928 & 0.631 & 0.794 & 1.089 & 0.873 & 0.753 & 0.668 & 0.585 \\
 0.4263 & 0.738 & 0.932 & 1.272 & 1.027 & 0.886 & 0.776 & 0.681 \\
 0.3650 & 0.746 & 0.952 & 1.243 & 1.032 & 0.905 & 0.781 & 0.683 \\
 0.3082 & 0.765 & 0.977 & 1.238 & 1.045 & 0.924 & 0.801 & 0.704 \\
 0.2554 & 0.534 & 0.649 & 0.801 & 0.692 & 0.607 & 0.530 & 0.469 \\
 0.2061 & 0.524 & 0.634 & 0.801 & 0.684 & 0.584 & 0.509 & 0.445 \\
 0.1599 & 0.645 & 0.773 & 1.008 & 0.845 & 0.702 & 0.606 & 0.520 \\
 0.1165 & 0.629 & 0.754 & 1.020 & 0.842 & 0.673 & 0.570 & 0.474 \\
 0.0756 & 0.508 & 0.564 & 0.774 & 0.640 & 0.506 & 0.431 & 0.360 \\
 0.0369 & 0.543 & 0.584 & 0.831 & 0.672 & 0.526 & 0.447 & 0.374 \\
 0.0000 & 0.580 & 0.630 & 0.895 & 0.728 & 0.571 & 0.483 & 0.404 \\
            \hline
         \end{array}
     $$ 
   \end{center}
   \end{table*}

   \begin{table*}
    \begin{center}
     \caption[Absorption corrections for the Kroupa model in the HST system]
     {Absorption corrections $m_{X,including abs.}-m_{X,abs.less}$ for the Kroupa model in the HST system.}
     \label{ACkroupahst}
    $$ 
     \begin{array}{rrrrrrrrrr}
      \hline
 redshift & U336 & B439 & V555 & V606 & R675 & I814 & J110 & H160 & K222 \\
            \hline
 9.5116 &     0 &     0 &     0 &     0 &     0 &     0 &     0 &     0 &     0 \\
 5.6177 & 0.267 & 0.199 & 0.151 & 0.143 & 0.139 & 0.107 & 0.100 & 0.100 & 0.065 \\
 4.0444 & 0.546 & 0.422 & 0.340 & 0.307 & 0.276 & 0.255 & 0.261 & 0.191 & 0.154 \\
 3.1576 & 0.627 & 0.553 & 0.459 & 0.435 & 0.413 & 0.436 & 0.396 & 0.283 & 0.222 \\
 2.5758 & 0.869 & 0.705 & 0.641 & 0.629 & 0.634 & 0.649 & 0.536 & 0.419 & 0.317 \\
 2.1589 & 1.024 & 0.862 & 0.824 & 0.841 & 0.887 & 0.770 & 0.641 & 0.516 & 0.391 \\
 1.8422 & 1.069 & 0.967 & 0.986 & 1.004 & 1.004 & 0.861 & 0.757 & 0.617 & 0.472 \\
\hline
         \end{array}
     $$ 
   \end{center}
   \end{table*}

   \begin{table*}
    \begin{center}
     \caption[Inclination corrections for the Kroupa model in the SDSS system]
     {Inclination corrections $m_{X,diagonal}-m_{X,faceon}$ for the Kroupa model in the SDSS system.}
     \label{ICkroupasdss}
    $$ 
     \begin{array}{rrrrrrrr}
      \hline
 redshift & BOL & V_j & u & g & r & i & z \\
            \hline
 9.5116 &      0 &      0 &      0 &      0 &      0 &      0 &      0 \\
 5.6177 &  0.037 &  0.065 &  0.079 &  0.049 &  0.064 &  0.055 &  0.047 \\
 4.0444 &  0.071 &  0.110 &  0.121 &  0.121 &  0.098 &  0.088 &  0.086 \\
 3.1576 &  0.119 &  0.160 &  0.211 &  0.174 &  0.152 &  0.151 &  0.163 \\
 2.5758 &  0.170 &  0.219 &  0.257 &  0.227 &  0.218 &  0.233 &  0.217 \\
 2.1589 &  0.243 &  0.329 &  0.378 &  0.336 &  0.337 &  0.335 &  0.290 \\
 1.8422 &  0.275 &  0.378 &  0.377 &  0.365 &  0.396 &  0.358 &  0.334 \\
\hline
         \end{array}
     $$ 
   \end{center}
   \end{table*}

   \begin{table*}
    \begin{center}
     \caption[Inclination corrections for the Kroupa model in the HST system]
     {Inclination corrections $m_{X,diagonal}-m_{X,faceon}$ for the Kroupa model in the HST system.}
     \label{ICkroupahst}
    $$ 
     \begin{array}{rrrrrrrrrr}
      \hline
 redshift & U336 & B439 & V555 & V606 & R675 & I814 & J110 & H160 & K222 \\
            \hline
 9.5116 &      0 &      0 &      0 &      0 &      0 &      0 &      0 &      0 &      0 \\
 5.6177 &  0.081 &  0.062 &  0.065 &  0.061 &  0.060 &  0.048 &  0.045 &  0.045 &  0.030 \\
 4.0444 &  0.128 &  0.137 &  0.110 &  0.101 &  0.093 &  0.087 &  0.088 &  0.067 &  0.054 \\
 3.1576 &  0.206 &  0.184 &  0.163 &  0.157 &  0.151 &  0.158 &  0.145 &  0.111 &  0.090 \\
 2.5758 &  0.267 &  0.234 &  0.222 &  0.220 &  0.222 &  0.226 &  0.198 &  0.167 &  0.134 \\
 2.1589 &  0.392 &  0.345 &  0.330 &  0.334 &  0.348 &  0.310 &  0.262 &  0.213 &  0.170 \\
 1.8422 &  0.379 &  0.363 &  0.374 &  0.382 &  0.384 &  0.344 &  0.313 &  0.269 &  0.218 \\
\hline
         \end{array}
     $$ 
   \end{center}
   \end{table*}

   \begin{table*}
    \begin{center}
     \caption[Inclination corrections for the Kroupa model in the SDSS system]
     {Inclination corrections $m_{X,edgeon}-m_{X,faceon}$ for the Kroupa model in the SDSS system.}
     \label{ICekroupasdss}
    $$ 
     \begin{array}{rrrrrrrr}
      \hline
 redshift & BOL & V_j & u & g & r & i & z \\
            \hline
 9.5116 &     0 &     0 &     0 &     0 &     0 &     0 &     0 \\
 5.6177 & 0.067 & 0.121 & 0.145 & 0.072 & 0.119 &   0.1 & 0.086 \\
 4.0444 & 0.150 & 0.228 & 0.203 & 0.248 & 0.207 & 0.189 & 0.185 \\
 3.1576 & 0.259 &  0.34 & 0.419 & 0.359 & 0.328 & 0.326 & 0.345 \\
 2.5758 & 0.342 & 0.426 & 0.478 & 0.438 & 0.426 & 0.447 & 0.425 \\
 2.1589 & 0.482 & 0.613 & 0.637 & 0.614 & 0.624 & 0.625 & 0.584 \\
 1.8422 & 0.576 & 0.746 & 0.746 & 0.729 & 0.767 & 0.723 & 0.693 \\
 1.5915 & 0.700 & 0.943 & 0.918 & 0.914 & 0.929 & 0.870 & 0.831 \\
 1.3868 & 0.791 & 1.098 & 1.063 & 1.088 & 1.034 & 0.969 & 0.911 \\
 1.2156 & 0.869 & 1.170 & 1.172 & 1.212 & 1.108 & 1.040 & 0.974 \\
 1.0696 & 0.930 & 1.165 & 1.185 & 1.226 & 1.131 & 1.078 & 1.049 \\
 0.9432 & 1.034 & 1.327 & 1.374 & 1.393 & 1.293 & 1.220 & 1.188 \\
 0.8322 & 1.004 & 1.318 & 1.382 & 1.362 & 1.290 & 1.236 & 1.207 \\
 0.7336 & 0.856 & 1.189 & 1.286 & 1.228 & 1.156 & 1.120 & 1.086 \\
 0.6453 & 0.894 & 1.226 & 1.385 & 1.282 & 1.179 & 1.141 & 1.095 \\
 0.5655 & 0.916 & 1.261 & 1.440 & 1.322 & 1.217 & 1.178 & 1.129 \\
 0.4928 & 0.900 & 1.252 & 1.375 & 1.296 & 1.229 & 1.197 & 1.150 \\
 0.4263 & 0.900 & 1.279 & 1.452 & 1.340 & 1.254 & 1.200 & 1.145 \\
 0.3650 & 0.943 & 1.360 & 1.603 & 1.438 & 1.325 & 1.249 & 1.180 \\
 0.3082 & 0.861 & 1.311 & 1.460 & 1.353 & 1.282 & 1.208 & 1.130 \\
 0.2554 & 0.902 & 1.338 & 1.506 & 1.381 & 1.298 & 1.216 & 1.129 \\
 0.2061 & 0.837 & 1.318 & 1.515 & 1.370 & 1.264 & 1.172 & 1.065 \\
 0.1599 & 0.757 & 1.293 & 1.490 & 1.352 & 1.224 & 1.107 & 0.960 \\
 0.1165 & 0.726 & 1.297 & 1.520 & 1.375 & 1.213 & 1.073 & 0.892 \\
 0.0756 & 0.687 & 1.229 & 1.482 & 1.329 & 1.136 & 0.975 & 0.772 \\
 0.0369 & 0.583 & 1.127 & 1.403 & 1.247 & 1.025 & 0.840 & 0.613 \\
 0.0000 & 0.501 & 0.957 & 1.294 & 1.107 & 0.848 & 0.649 & 0.432 \\
            \hline
         \end{array}
     $$ 
   \end{center}
   \end{table*}

   \begin{table*}
    \begin{center}
     \caption[Inclination corrections for the Kroupa model in the HST system]
     {Inclination corrections $m_{X,diagonal}-m_{X,edgeon}$ for the Kroupa model in the HST system.}
     \label{ICekroupahst}
    $$ 
     \begin{array}{rrrrrrrrrr}
      \hline
 redshift & U336 & B439 & V555 & V606 & R675 & I814 & J110 & H160 & K222 \\
            \hline
 9.5116 &     0 &     0 &     0 &     0 &     0 &     0 &     0 &     0 &     0 \\
 5.6177 & 0.150 & 0.105 & 0.121 & 0.114 & 0.111 & 0.088 & 0.083 & 0.083 & 0.055 \\
 4.0444 & 0.217 & 0.289 & 0.229 & 0.213 & 0.198 & 0.186 & 0.187 & 0.145 & 0.119 \\
 3.1576 & 0.407 & 0.372 & 0.344 & 0.335 & 0.326 & 0.336 & 0.317 & 0.257 & 0.214 \\
 2.5758 & 0.493 & 0.447 & 0.431 & 0.428 & 0.432 & 0.438 & 0.400 & 0.356 & 0.303 \\
 2.1589 & 0.645 & 0.622 & 0.611 & 0.618 & 0.636 & 0.602 & 0.551 & 0.492 & 0.426 \\
 1.8422 & 0.745 & 0.724 & 0.741 & 0.751 & 0.752 & 0.705 & 0.661 & 0.601 & 0.525 \\
\hline
         \end{array}
     $$ 
   \end{center}
   \end{table*}

\section{Other quantities}

   \begin{table*}
    \begin{center}
     \caption[Petrosian radii for the Kroupa model]
     {Petrosion radii [kpc] for the Kroupa model.}
     \label{R90kroupa}
    $$ 
     \begin{array}{rrrrrrrr}
      \hline
 t [Gyr] & petroR90_{BOL} & petroR90_{Vj} & petroR90_u & petroR90_g & petroR90_r & petroR90_i & petroR90_z \\
            \hline
  0.5 & 16.69 & 16.72 & 16.71 & 16.72 & 16.73 & 16.62 & 15.56 \\
  1.0 & 3.499 & 3.348 & 3.435 & 3.389 & 3.312 & 3.256 & 3.217 \\
  1.5 & 4.467 & 4.353 & 4.451 & 4.410 & 4.306 & 4.233 & 4.158 \\
  2.0 & 5.704 & 5.374 & 5.618 & 5.474 & 5.298 & 5.195 & 5.104 \\
  2.5 & 6.312 & 6.184 & 6.339 & 6.270 & 6.118 & 6.011 & 5.896 \\
  3.0 & 6.778 & 6.661 & 6.821 & 6.759 & 6.582 & 6.454 & 6.323 \\
  3.5 & 7.465 & 7.150 & 7.468 & 7.299 & 7.040 & 6.881 & 6.720 \\
  4.0 & 7.504 & 7.424 & 7.654 & 7.553 & 7.326 & 7.160 & 6.969 \\
  4.5 & 7.596 & 7.577 & 7.772 & 7.705 & 7.473 & 7.301 & 7.102 \\
  5.0 & 7.860 & 7.767 & 8.076 & 7.938 & 7.646 & 7.472 & 7.246 \\
  5.5 & 7.946 & 7.902 & 8.212 & 8.066 & 7.781 & 7.616 & 7.408 \\
  6.0 & 8.282 & 8.234 & 8.459 & 8.369 & 8.132 & 7.980 & 7.805 \\
  6.5 & 8.611 & 8.666 & 8.862 & 8.797 & 8.561 & 8.406 & 8.197 \\
  7.0 & 8.729 & 8.885 & 9.001 & 8.999 & 8.785 & 8.637 & 8.447 \\
  7.5 & 8.992 & 9.094 & 9.261 & 9.222 & 8.989 & 8.821 & 8.620 \\
  8.0 & 8.907 & 9.092 & 9.226 & 9.211 & 8.992 & 8.846 & 8.647 \\
  8.5 & 9.007 & 9.096 & 9.387 & 9.237 & 8.986 & 8.797 & 8.598 \\
  9.0 & 8.993 & 9.163 & 9.360 & 9.284 & 9.067 & 8.916 & 8.710 \\
  9.5 & 8.830 & 9.052 & 9.176 & 9.164 & 8.966 & 8.843 & 8.677 \\
 10.0 & 8.893 & 9.095 & 9.310 & 9.218 & 9.002 & 8.856 & 8.671 \\
 10.5 & 8.805 & 9.048 & 9.283 & 9.181 & 8.950 & 8.809 & 8.641 \\
 11.0 & 8.778 & 9.029 & 9.283 & 9.157 & 8.937 & 8.804 & 8.641 \\
 11.5 & 8.596 & 8.903 & 9.081 & 9.014 & 8.820 & 8.702 & 8.541 \\
 12.0 & 8.395 & 8.740 & 8.827 & 8.833 & 8.677 & 8.582 & 8.444 \\
 12.5 & 8.445 & 8.756 & 8.928 & 8.858 & 8.680 & 8.571 & 8.419 \\
 13.0 & 8.304 & 8.671 & 8.901 & 8.776 & 8.601 & 8.500 & 8.359 \\
 13.5 & 8.133 & 8.477 & 8.717 & 8.586 & 8.407 & 8.310 & 8.174 \\
            \hline
         \end{array}
     $$ 
   \end{center}
   \end{table*}

   \begin{table*}
    \begin{center}
     \caption[Petrosian radii for the Kroupa model]
     {Petrosion radii [arcsec] for the Kroupa model.}
     \label{r90kroupa}
    $$ 
     \begin{array}{rrrrrrrr}
      \hline
 redshift & petroR90_{BOL} & petroR90_{Vj} & petroR90_u & petroR90_g & petroR90_r & petroR90_i & petroR90_z \\
            \hline
 9.5116 & 3.886 &     0 &     0 &     0 &     0 &     0 & 3.879 \\
 5.6177 & 0.585 & 0.610 & 0.673 & 1.038 & 0.615 & 0.604 & 0.596 \\
 4.0444 & 0.643 & 0.665 & 0.865 & 0.660 & 0.655 & 0.654 & 0.650 \\
 3.1576 & 0.748 & 0.794 & 0.808 & 0.808 & 0.790 & 0.778 & 0.785 \\
 2.5758 & 0.785 & 0.809 & 0.817 & 0.809 & 0.807 & 0.816 & 0.807 \\
 2.1589 & 0.816 & 0.836 & 0.849 & 0.838 & 0.839 & 0.840 & 0.828 \\
 1.8422 & 0.883 & 0.925 & 0.941 & 0.929 & 0.928 & 0.909 & 0.892 \\
\hline
         \end{array}
     $$ 
   \end{center}
   \end{table*}

   \begin{table*}
    \begin{center}
     \caption[Concentration parameters for the Kroupa model]
     {Concentration parameters for the Kroupa model.}
     \label{Ckroupa}
    $$ 
     \begin{array}{rrrrrrrr}
      \hline
 redshift & C_{BOL} & C_{Vj} & C_u & C_g & C_r & C_i & C_z \\
            \hline
 9.5116 &     - &     - &     - &     - &     - &     - &     - \\
 5.6177 &     - &     - &     - &     - &     - &     - &     - \\
 4.0444 &     - &     - &     - &     - &     - &     - &     - \\
 3.1576 &     - &     - &     - &     - &     - &     - &     - \\
 2.5758 &     - &     - &     - &     - &     - &     - &     - \\
 2.1589 &     - &     - &     - &     - &     - &     - &     - \\
 1.8422 &     - &     - &     - &     - &     - &     - &     - \\
 1.5915 & 0.234 & 0.298 & 0.337 & 0.311 & 0.292 & 0.264 & 0.237 \\
 1.3868 & 0.135 & 0.135 & 0.135 & 0.135 & 0.135 & 0.135 & 0.134 \\
 1.2156 & 0.229 & 0.228 & 0.221 & 0.230 & 0.226 & 0.224 & 0.220 \\
 1.0696 & 0.290 & 0.270 & 0.265 & 0.277 & 0.268 & 0.264 & 0.265 \\
 0.9432 & 0.378 & 0.363 & 0.360 & 0.370 & 0.361 & 0.352 & 0.353 \\
 0.8322 & 0.398 & 0.383 & 0.382 & 0.384 & 0.383 & 0.378 & 0.379 \\
 0.7336 & 0.297 & 0.264 & 0.243 & 0.255 & 0.266 & 0.267 & 0.274 \\
 0.6453 & 0.284 & 0.240 & 0.232 & 0.236 & 0.238 & 0.244 & 0.250 \\
 0.5655 & 0.311 & 0.269 & 0.261 & 0.268 & 0.264 & 0.270 & 0.278 \\
 0.4928 & 0.319 & 0.275 & 0.258 & 0.274 & 0.274 & 0.283 & 0.291 \\
 0.4263 & 0.324 & 0.271 & 0.259 & 0.271 & 0.273 & 0.282 & 0.291 \\
 0.3650 & 0.353 & 0.303 & 0.322 & 0.313 & 0.303 & 0.306 & 0.311 \\
 0.3082 & 0.371 & 0.314 & 0.329 & 0.320 & 0.315 & 0.319 & 0.323 \\
 0.2554 & 0.387 & 0.333 & 0.338 & 0.334 & 0.334 & 0.339 & 0.344 \\
 0.2061 & 0.410 & 0.343 & 0.362 & 0.346 & 0.343 & 0.347 & 0.352 \\
 0.1599 & 0.441 & 0.373 & 0.399 & 0.378 & 0.373 & 0.375 & 0.382 \\
 0.1165 & 0.456 & 0.397 & 0.434 & 0.405 & 0.394 & 0.393 & 0.401 \\
 0.0756 & 0.492 & 0.417 & 0.463 & 0.425 & 0.415 & 0.416 & 0.427 \\
 0.0369 & 0.517 & 0.438 & 0.470 & 0.443 & 0.438 & 0.441 & 0.455 \\
 0.0000 & 0.556 & 0.479 & 0.512 & 0.484 & 0.478 & 0.481 & 0.497 \\
            \hline
         \end{array}
     $$ 
   \end{center}
   \end{table*}

\end{document}